\documentclass[aps,pre,10pt,twocolumn,superscriptaddress,byrevtex,bibnotes,showpacs,longbibliography]{revtex4-1}

\usepackage{amsmath,amssymb,wasysym,latexsym}
\usepackage{graphicx,subfigure}
\usepackage[section]{placeins}
\usepackage{color}
\usepackage{microtype}
\usepackage{relsize}
\usepackage[colorlinks=true,citecolor=darkgreen,linkcolor=blue,urlcolor=blue]{hyperref}
\definecolor{darkgreen}{rgb}{0.0, 0.4, 0.0}

\newcommand{\mb}[1]{ \boldsymbol #1 }
\newcommand{\ti}[1]{ \tilde{#1} }
\newcommand{\ra}{\rightarrow}
\newcommand{\E}{\mathbb{E}}

\begin{document}

\title{Local structure of current fluctuations in diffusive systems beyond one dimension}
\author{Rodrigo Villavicencio-Sanchez}
\email{rvillavicenciosanchez@unisa.it}
\affiliation{School of Mathematical Sciences, Queen Mary University of London, London E1 4NS, UK}
\affiliation{Dipartimento di Fisica ``E.~R. Caianiello'', Universit\`a di Salerno, via Giovanni Paolo II 132, 84084 Fisciano (SA), Italy.}
\author{Rosemary J. Harris}
\email{rosemary.harris@qmul.ac.uk}
\affiliation{School of Mathematical Sciences, Queen Mary University of London, London E1 4NS, UK}
\date{\today}

\begin{abstract}
In order to illuminate the properties of current fluctuations in more than one dimension, we use a lattice-based Markov process driven into a non-equilibrium steady state. Specifically, we perform a detailed study of the particle current fluctuations in a two-dimensional zero-range process with open boundary conditions and probe the influence of the underlying geometry by comparing results from a square and a triangular lattice. Moreover, we examine the structure of local currents corresponding to a given global current fluctuation and comment on the role of spatial inhomogeneities for the discrepancies observed in testing some recent fluctuation symmetries.
\end{abstract}

\pacs{02.50.-r, 05.40.-a, 05.70.Ln}
\maketitle

\section{Introduction}

The understanding of non-equilibrium physics is of great relevance for many scenarios ranging from granular materials, chemical reactions and molecular motors to traffic jams~\cite{puglisi2005a,qian2006,julicher1997,chowdhury2000}. In particular, much interest is directed towards the study of the probability of rare events or trajectories in stochastic models. This has led to the establishment of fluctuation theorems, which are some of the most general results for systems out of equilibrium (for reviews see~\cite{ritort2003,harris2007,seifert2012}). Within the stochastic framework, interacting particle systems have enjoyed widespread use to model non-equilibrium steady states (NESSs). Most such models are one-dimensional; we expect a richer phenomenology in more than one dimension just as higher dimensional equilibrium systems are qualitatively different from their one-dimensional counterparts. 

The study of NESSs in more than one dimension has led to the recent discovery of symmetries for global current fluctuations in macroscopic systems~\cite{hurtado2011,villavicencio2014}. In particular, by considering a diffusive lattice gas on a $d$-dimensional (hyper-cubic) lattice of side length $L$, fluctuation relations were obtained for a time-averaged global current defined as
\begin{equation}
\mb{J}=\frac{1}{t}\int_{0}^{t}\mathrm{d}\tau \int_{\Omega}\mathrm{d}\mb{r}\,\mb{j}(\mb{r},\tau). \label{global-current}
\end{equation}
Here, the local current, $\mb{j}(\mb{r},t)$, is assumed to obey the continuity equation, and a diffusive scaling is applied. Namely, space is scaled to $\Omega=[0,1]^{d}$, and time is scaled by $1/L^2$. Then, according to the macroscopic fluctuation theory (MFT) the probability of observing a rare global current can be calculated from the minimization of an action functional that depends on the local value of the current and density~\cite{bertini2006,bertini2015}. Physically, this means that out of all the possible ways to generate a fluctuation, the overwhelmingly most likely to be realized corresponds to a specific optimal density profile (ODP) and optimal current profile (OCP). Under some hypotheses, notably a spatially homogeneous OCP, it is possible to manipulate the MFT action functional to obtain that global current fluctuations satisfy the relation
\begin{equation}
\lim\limits_{tL^d\ra\infty}-\frac{1}{tL^d}\ln\frac{P(\mb{J}',t)}{P(\mb{J},t)}=\mb{E}\cdot\left(\mb{J}-\mb{J}'\right) \label{general-fluctuation-relation}
\end{equation}
for isometric fluctuations such that 
\begin{equation}
|\mb{J}|=|\mb{J}'|. \label{ellipse-equation}
\end{equation} 
Here, $P(\mb{J},t)$ is the probability of observing the global current fluctuation $\mb{J}$ during a time interval $t$, $\mb{E}$ is a constant that depends on the boundary and the bulk driving of the system, and $|\cdot|$ denotes the modulus. Note that this isometric fluctuation relation (IFR) reduces to the renowned Gallavotti-Cohen-type fluctuation symmetry~\cite{evans1993,gallavotti1995,kurchan1998,lebowitz1999} for anti-parallel currents but it also relates in a surprisingly simple manner currents in different directions~\cite{hurtado2011}. Furthermore, in~\cite{villavicencio2014} the IFR was generalized to anisotropic systems, where some discrepancies were also noted between the global current fluctuations predicted to satisfy the symmetry at a macroscopic level and those in models on (large) finite size lattices. Remarkably, these fluctuation relations have recently been tested experimentally as reported in~\cite{kumar2015}, where fluctuations of the velocity of a tapered rod are shown to be well approximated by the anisotropic generalization of the IFR.

In this paper, we use a continuous-time Markov process driven by the boundaries into an NESS in order to study the detailed properties of current fluctuations in more than one dimension. In particular, a two-dimensional zero-range process (ZRP) is solved to study: \emph{a}) the influence of the underlying lattice geometry on the probability of a global current (and density) fluctuation, and \emph{b}) the most likely local current structure of the OCP associated to a given global current fluctuation. Specifically, we test whether the hypothesis of homogeneous OCP may have to be adjusted for some systems with open boundary conditions, explaining the above-mentioned discrepancies observed for rare global currents. 

The paper is structured as follows. In Section~\ref{s:maccurrfluc} we introduce some definitions from large deviation theory commonly used in the study of NESSs. In Section~\ref{s:micapp} we explain the so-called quantum Hamiltonian formalism which we employ to study the stationary state and the probability of measuring rare particle current fluctuations. In Section~\ref{s:sqrtrilatt} we solve exactly an anisotropic two-dimensional (2-$d$) ZRP on square and triangular lattices, allowing us to analyse fluctuations and compare the effect of the underlying geometry as the system size increases. In Section~\ref{s:inhomogcurr} we refine our calculations to show that local current fluctuations have a more complex structure than implicitly assumed in some other works, and highlight the relevance of our results for the anisotropic fluctuation relation (AFR) and the original IFR. In Section~\ref{s:outlook} we summarize our findings and comment on some remaining open questions. In addition, we include various technical details in a series of appendices.

Note that some of these results were already presented in a briefer work~\cite{villavicencio2014}; significantly, we here offer convincing evidence (Section~\ref{s:inhomogcurr}) to support the conjecture made there on the local structure of current fluctuations. In addition, we present a new extension to the triangular lattice (Section~\ref{s:sqrtrilatt}) and provide many previously unpublished calculational details including a more explicit derivation of the AFR than the one shown in~\cite{villavicencio2014} (Appendix~\ref{a:alternativeafr}) as well as the corresponding macroscopic optimization argument for the 2-$d$ ZRP with non-decreasing interactions (Appendix~\ref{a:mft-calculations}).


\section{Large deviation theory: current fluctuations}\label{s:maccurrfluc}

One of the main goals in this paper is to study the structure of the current profiles that yield a particular global current fluctuation. However, our results are directly related to the accuracy of the IFR and the AFR for open systems. In this section we introduce these two fluctuation relations beginning from the framework of large deviation theory.

The study of NESSs involves understanding the probability of measuring rare currents. In a lattice-gas model for example, a current is understood as the net number of particles that jump between two adjacent sites in a positive direction (arbitrarily chosen) during a given time window. When a system is in an NESS the mean flux of particles is, generally, a constant different from zero. Moreover, it is known that in many cases such currents obey a large deviation principle (LDP), for instance the global current $\mb{J}$ in Eq.~(\ref{global-current}) satisfies
\begin{equation}
\hat{e}(\mb{J})=\lim\limits_{tL^d\ra\infty}-\frac{1}{tL^d}\ln P(\mb{J},t), \label{ldp}
\end{equation}
where $\hat{e}(\mb{J})$ is a rate function (RF) encoding the probability, $P(\mb{J},t)$, of observing a given current in the long-time limit~\cite{touchette2009,dembo1998}. 

In order to calculate the RF, it is useful to compute first the scaled cumulant generating function (SCGF) 
\begin{equation}
e(\mb{\lambda})=\lim\limits_{tL^d\ra\infty}-\frac{1}{tL^d}\ln\langle \exp\left(-tL^d\mb{\lambda} \cdot\mb{J}\right) \rangle \label{scgf}
\end{equation}
where $\mb{\lambda}$ is a vector conjugate to $\mb{J}$, and $\langle\cdot\rangle$ denotes an expectation value. It is well known that when the SCGF is differentiable we can compute the RF using the G\"artner-Ellis Theorem, which relates these functions via the Legendre transform~\cite{touchette2009}
\begin{equation}
\hat{e}(\mb{J})=\max\limits_{\mb{\lambda}}\left\{e(\mb{\lambda})-\mb{\lambda}\cdot\mb{J}\right\}.\label{legetransf}
\end{equation}
As we will see below, much can be learned from the SCGF about the probability of the currents, but in the rest of this section we remind the reader about some fluctuation relations which will be discussed later in the paper.

At this point it is possible to use the LDP~(\ref{ldp}) to rewrite the fluctuation relation~(\ref{general-fluctuation-relation}) in terms of the RF as
\begin{equation}
\hat{e}(\mb{J})-\hat{e}(\mb{J}')=\mb{E}\cdot(\mb{J}'-\mb{J}), \label{afr-rf}
\end{equation}
for global currents satisfying $|\mb{J}|=|\mb{J}'|$. Here, the constant $\mb{E}$ can be seen as an external field driving the system out of equilibrium. Moreover, this also implies a symmetry at the level of the SCGF which is expressed simply as
\begin{equation}
e(\mb{\lambda})=e(\mb{\lambda}'), \label{SCGF-FR}
\end{equation}
for values of $\mb{\lambda}$ such that
\begin{equation}
|\mb{\lambda}-\mb{E}| = |\mb{\lambda}'-\mb{E}|. \label{ellipse-equation-lambdas}
\end{equation}
Here we note that Eq.~(\ref{ellipse-equation-lambdas}) also corresponds to the equation of a circle for the conjugate parameter of the current, but with centre at the constant field $\mb{E}$. As mentioned above, such a symmetry yields as a special case the Gallavotti-Cohen-type relation for forward and backward currents. The AFR (derived in Appendix~\ref{a:alternativeafr}) is a generalization of Eqs.~(\ref{afr-rf})--(\ref{ellipse-equation-lambdas}) where ellipses, instead of circles, relate current fluctuations in different directions. 

In the following section, we will explain how to study fluctuations of a similar space-integrated global current, in finite (microscopic) ZRPs. Later, we will explain how to scale such a current to compare the results with the ones obtained from a macroscopic point of view.

\section{Zero-range process: microscopic approach}\label{s:micapp}

\subsection{Definition of the model}

\begin{figure*}[t]
	\centering
	\subfigure[][]{\includegraphics[width=3.4in]{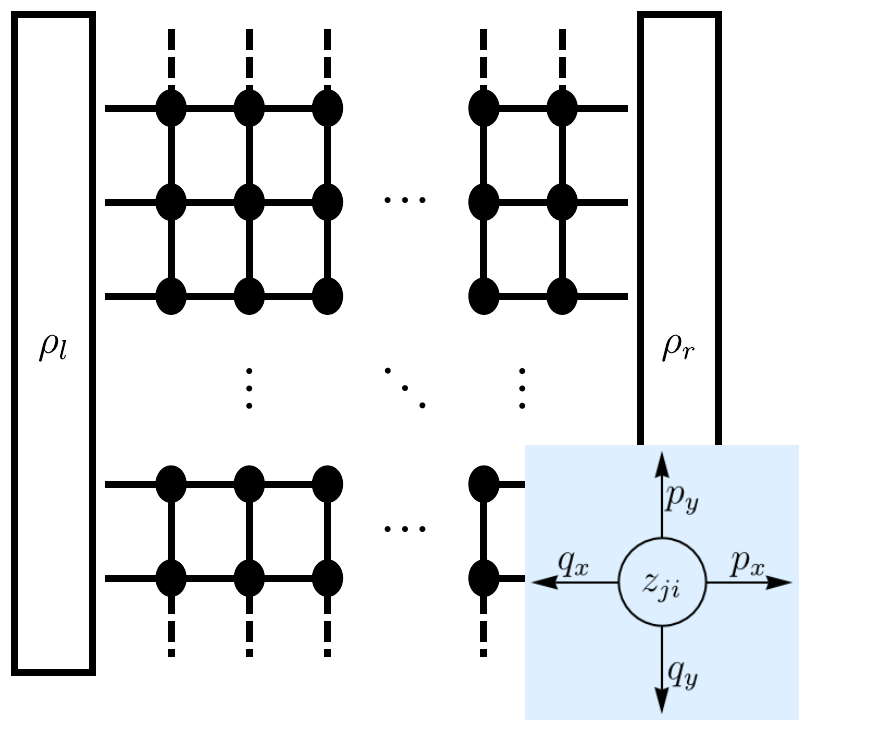}	\label{lattices-left}}
		\quad
	\subfigure[][]{\includegraphics[width=3.4in]{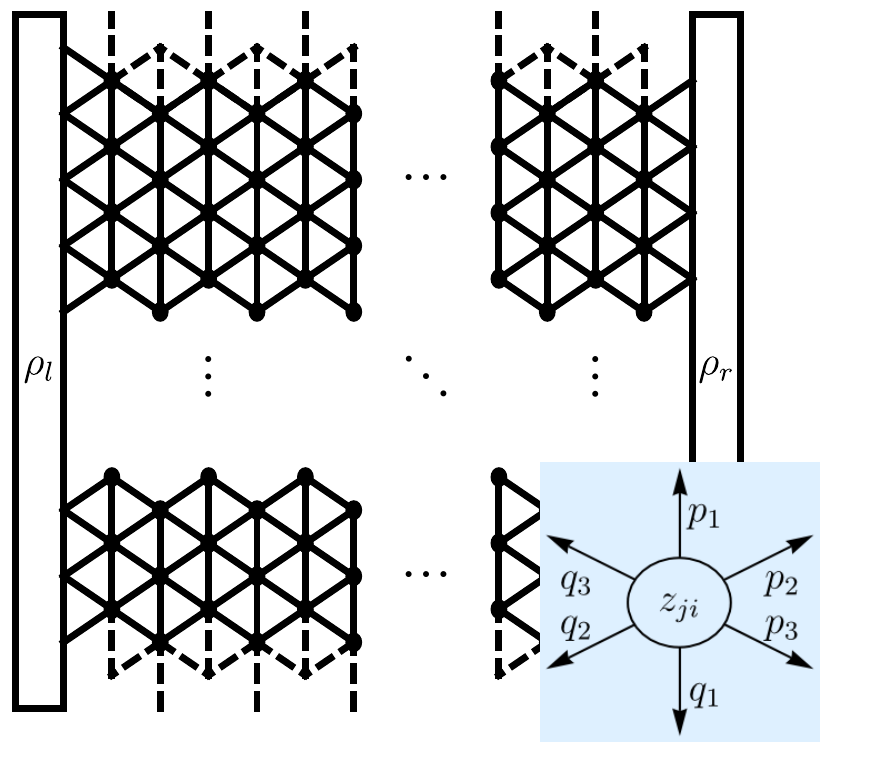}	\label{lattices-right}}
	\caption[Square and triangular 2-d lattices]{(Color online) Different underlying geometries, \subref{lattices-left}~square lattice and \subref{lattices-right}~triangular lattice, with periodic boundary conditions in the vertical direction (\emph{y}-direction) and open boundary conditions in the horizontal direction ($x$-direction). Sites are identified by their row ($y$-coordinate, first subindex) and column ($x$-coordinate, second subindex) as shown in the insets. In the triangular geometry, sites on the same row are joined by a zigzag line. In the microscopic model open boundaries are specified by the input rates ($\alpha$ and $\delta$) as well as the output rates ($\gamma$ and $\beta$) for left and right boundaries respectively (see also Fig.~\ref{bdryrates}); the input rates can be related to the left and right reservoir densities $\rho_l$ and $\rho_r$.}\label{lattices}
\end{figure*}

The ZRP is a model of interacting particles introduced in 1970~\cite{spitzer1970} and since studied on general lattices~\cite{andjel1982,evans2005b}. Particles are allowed to accumulate up to any non-negative number on each site of the lattice (e.g., Fig.~\ref{lattices}). The top-most particle of each site jumps to a neighbouring site after an exponentially distributed waiting time, where the hopping rate is proportional to an on-site particle interaction factor, $w_n$. As the name of the model suggests, $w_n$ depends exclusively on the total occupation of the departure site. Indeed, such an interaction can cause a phase transition where a macroscopic proportion of particles in the system accumulates on a single site of the lattice~\cite{evans2000,levine2005}. Similar condensation phenomena are of wide interest in connection with granular systems~\cite{shim2004} and wealth models~\cite{burda2002} among other topics.

In order to study this zero-range model we employ a general framework~\cite{schutz2001}, referred to as the quantum Hamiltonian formalism, in which the master equation of the system is written in a form resembling a Schr\"odinger equation. Within this approach one can compute the probability of particle configurations in the system, as well as other important quantities, such as the time-averaged particle current. 

We begin by defining the configuration, $\mb{n}=\left(n_1,n_2,...,n_N\right)$, containing the number of particles on each of the $N$ sites of the lattice at a given time. Then, each configuration $\mb{n}$ is associated with an element of a basis, $|n\rangle$, to construct the probability vector 
\begin{equation}
	|P\rangle=\sum\limits_{\mb{n}} P(\mb{n})|n\rangle,\label{probvec}
\end{equation}
where $P(\mb{n})$ is the probability of finding the system in configuration $\mb{n}$. 

The time evolution of the probability vector is described by the master equation
\begin{equation}
\frac{d|P\rangle}{dt}=-H|P\rangle.\label{mastereq}
\end{equation}
Here the stochastic generator, $H$, also called the Hamiltonian, contains the hopping rates between all states of the system, and can be written in terms of the ladder operators
\begin{equation}
 a_i^+=\left( \begin{array}{cccc}
                0 & 0 & 0 & \quad \\
		1 & 0 & 0 & \dots \\
		0 & 1 & 0 & \quad \\
		\quad & \vdots & \quad & \ddots
              \end{array} \right)
\text{ and }a_i^-=\left( \begin{array}{cccc}
                0 & w_1 & 0 & \quad \\
		0 & 0 & w_2 & \dots \\
		0 & 0 & 0 & \quad \\
		\quad & \vdots & \quad & \ddots
              \end{array} \right), \label{ladderops}
\end{equation}
which act exclusively on the $i^{th}$ component of the configuration vector. 

On each lattice shown in Fig.~\ref{lattices}, particles from the bulk jump with rates $p_k$ and $q_k$ in the positive or negative $k$-direction (respectively), as indicated by the insets. Thus, a particle jump is represented in the quantum Hamiltonian formalism by the simultaneous annihilation and creation of one particle, at the departure and target sites respectively, with the operators~(\ref{ladderops}) times the corresponding hopping rate. Furthermore, particles are injected at constant rates $\alpha_k$ and $\delta_k$, or extracted with rates $\gamma_k$ and $\beta_k$, both at the left and right boundaries respectively. 

Note that for the two dimensional systems we study, it is convenient to identify sites, and corresponding ladder operators, with two subindices as done in Appendix~\ref{a:hamiltonians} where we explicitly show the Hamiltonians corresponding to the square and triangular lattices (Eqs.~(\ref{Hs}) and~(\ref{Ht}) respectively). We now turn to study the time-independent solution of~(\ref{mastereq}), i.e., the steady state.

\subsection{Steady state solution}

Typically, to drive a system out of equilibrium, we let it interact with more than one reservoir and expect it to reach an NESS in the long-time limit. In the present context, by considering a ZRP with open boundary conditions where the input and output rates are different at the two borders we expect the system to approach a time-independent solution with constant mean current through the system. This means that the left-hand side (LHS) of~(\ref{mastereq}) vanishes, leaving 	us with the eigenvalue equation
\begin{equation}
H|P^*\rangle=0\label{ness}
\end{equation}
implying the stationary state, $|P^*\rangle$, is the eigenvector with eigenvalue zero. Similarly to certain other interacting particle models, for the ZRP with open boundaries the vector $|P^*\rangle$ is given by the product measure~\cite{levine2005,evans2006}
\begin{equation}
|P^*\rangle=|P_1^*\rangle\otimes|P_2^*\rangle\otimes...\otimes|P_N^*\rangle, \label{tensorprod}
\end{equation}
i.e., the probability distribution factorizes over the sites. For the ZRP, it can be shown that the marginal for the $i^{\text{th}}$ site is
\begin{equation}
	|P_i^*\rangle=\sum\limits_{n_i} P^*_i(n_i)|n_i\rangle, \label{siteprob}
\end{equation}
where the probability of finding $n_i$ particles on site $i$ is given by~\cite{levine2005}
\begin{equation}
P_i^*(n_i)=Z_i^{-1}z_i^{n_i}\prod\limits_{j=1}^{n_i}w_j^{-1}. \label{nparticsprob}
\end{equation}
Here $z_i$ is the fugacity of site \emph{i} and $Z_i$ is the grand canonical partition function
\begin{equation}
	Z_i=\sum\limits_{j=0}^{\infty}z_i^j\prod\limits_{n=1}^{j}w_n^{-1}. \label{grandcanpartfunc}
\end{equation}
Note that for some choices of the interaction term $w_n$ the radius of convergence, $z_{\max}$, of the sum~(\ref{grandcanpartfunc}) may be finite. Within the range of values where the partition function is well defined, the site densities are related to the fugacities via the equation
\begin{equation}
\rho_i=z_i\frac{\partial \ln Z_i}{\partial z_i}. \label{grandcanrel-fugdens}
\end{equation} 
Outside this range, the diverging $Z_i$ corresponds to the accumulation (condensation) of infinitely many particles on site \emph{i}. Here we aim to study current fluctuations within the fluid regime of the ZRP (i.e. no condensation), for this purpose, it is sufficient to consider $w_n$ as an increasing function of the number of particles. As in the one-dimensional ZRP with open boundaries~\cite{levine2005}, it turns out that the fugacities are independent of $w_n$. However, Eq.~(\ref{grandcanrel-fugdens}) will become relevant when we study the density profile related to a given current fluctuation~\footnote{In fact, interactions which allow a phase transition could be considered, but our analysis would be valid only in the fluid regime of the system. The domain of validity is determined by the radius of convergence of $Z$}.

In practice, to compute the fugacities we note that the creation and annihilation operators affect only the corresponding site component of the stationary state eigenvector according to
\begin{eqnarray}
a_i^+|P_i^*\rangle &=& z_i^{-1}d_i|P_i^*\rangle\label{upright}\\
a_i^-|P_i^*\rangle &=& z_i|P_i^*\rangle\label{downright}.
\end{eqnarray}
Here, we have defined the diagonal matrices $d_i$ with elements $d_{jk}=w_j\delta_{j,k}$ where $w_0=0$ by definition and $\delta_{j,k}$ is the Kronecker delta. Then for a lattice of $N$ sites, Eq.~(\ref{ness}) can be reduced using~(\ref{upright}) and~(\ref{downright}) to a system of $N$ coupled equations for the fugacities of the system. Note that in this framework, conservation of probability implies that the corresponding left eigenvector has every component equal to unity; we denote such an eigenvector by $\langle 1|$.

\subsection{Current fluctuations}

In addition to the probabilities of configurations, it is also possible to study the probabilities of fluctuations of particle currents within the system. Specifically, our goal is to quantify the probability that the time-averaged number of particle jumps between nearest neighbouring sites, in the whole lattice or a subset of it, attains a given rare value. This means that we have to modify the quantum Hamiltonian to count the number of particles that jump within the lattice during the observation time interval. 

To avoid confusion with the macroscopic current, we will use the variable $I$ to denote the space-integrated microscopic current measured across the lattice. Later we will clarify how to rescale this current to compare it with the macroscopic approach, but first we explain how the current is constructed. We define a certain time evolution of the system as the set of configurations, $\{\sigma\}=\{\sigma_1,\sigma_2,...,\sigma_\tau\}$, visited by the system during the time interval $[0,t]$. In one dimension it is clear that the net number of particle jumps is counted with an antisymmetric function; we let $\Theta_{\sigma_{i+1},\sigma_{i}}$ take the value $+1$ when particles jump forwards and $-1$ when particles jump backwards anywhere in the lattice (for a more general case see e.g.~\cite{rakos2008}). This way, the space-integrated current in one dimension is defined as the sum 
\begin{equation}
I(t,\{\sigma\})=\frac{1}{t}\sum\limits_{i=0}^{\tau-1}\Theta_{\sigma_{i+1},\sigma_{i}}. \label{spc-int-curr}
\end{equation}
In higher dimensional lattices, we will be interested in a similar vectorial variable with component $I_k(t,\{\sigma\})$ to count the number of jumps in the $k$-direction. For now we keep the one dimensional notation in order to demonstrate the framework.

In analogy to the macroscopic case, to compute the microscopic RF
\begin{equation}
\hat{e}_L(I)=\lim\limits_{t\ra\infty}-\frac{1}{t}\ln P(I,t), \label{micro-rf}
\end{equation}
it is useful to define first the SCGF
\begin{equation}
e_L(\lambda)=\lim\limits_{t\ra\infty}-\frac{1}{t}\ln\langle \exp\left(-t \lambda I\right)\rangle. \label{scgf-micro-1dim}
\end{equation}
It can be shown that the average on the right-hand side (RHS) of this relation can be written as $\langle\exp(-t\hat{H})\rangle$, where $\hat{H}$ is a modified Hamiltonian. In order to obtain $\hat{H}$, we have to multiply the terms of $H$ corresponding to particle transitions by $\exp(-\lambda)$ for jumps in the positive direction and by $\exp(\lambda)$ for jumps in the negative direction~\cite{harris2005}.

In cases when the eigenvalue spectrum of $\hat{H}$ is gapped the calculation of the SCGF can be done by noticing that in the long-time limit the exponential of the lowest eigenvalue, $\zeta(\lambda)$, dominates the average~(\ref{scgf-micro-1dim}). This leads to the result
\begin{equation}
e_L(\lambda)=\zeta(\lambda), \label{scgf-evalue}
\end{equation}
where we have assumed that the prefactors arising from the eigenstate decomposition are finite. Breakdown of this condition signifies condensation.

Furthermore, the right ground-state eigenvector, $|\psi\rangle$, turns out to have the same form as the stationary state~(\ref{tensorprod})--(\ref{grandcanpartfunc}) but with some modified fugacities, $\hat{z}_{i}(\lambda)$. In principle it is possible to calculate exactly the modified fugacities $|\psi_i\rangle$ by using relations analogous to Eq.~(\ref{upright}) and Eq.~(\ref{downright}), allowing us to determine also the SCGF. Notice that the lowest eigenvalue does not vanish in general. Indeed, one can verify consistency with the stationary state by substituting $\lambda=0$, for which the eigenvalue does become zero.

Finally, the RF is calculated via a Legendre transform similar to Eq.~(\ref{legetransf}) for microscopic currents. We remark that when the transform cannot be computed analytically, we can use the implicit relations
\begin{equation}
\begin{aligned}
I &= \frac{d e_L(\lambda))}{d \lambda} \\
\lambda &= -\frac{d \hat{e}_L(I)}{d I}
\end{aligned}\label{eq8}
\end{equation}
to calculate it numerically.

To obtain the density profile which gives rise to these currents, we have to compute the mean local occupation $\langle n_i \rangle$ taking into account the dynamics of the modified Hamiltonian. To do this, we need both left and right modified eigenvectors corresponding to the ground state eigenvalue. The left (row) eigenvector, $\langle \psi|$, is again a product with terms denoted by $\langle\psi_i|=(1,\ti{z}_i,\ti{z}_i^2,...)$. To calculate the left fugacities, $\ti{z}_i(\lambda)$, we use the modified Hamiltonian on $\langle\psi|$, where the ladder operators act according to the relations
\begin{eqnarray}
\langle \psi_i|a_i^+ &=& \langle \psi_i|\ti{z}_i\label{upleft}\\
\langle \psi_i|a_i^- &=& \langle \psi_i|\ti{z}_i^{-1}d_i.\label{downleft}
\end{eqnarray}
Here the dependence on $\lambda$ is left implicit; this is done from now on for both $\ti{z}_i$ and $\hat{z}_i$.

Using Eqs.~(\ref{upleft}) and~(\ref{downleft}), the left fugacities are obtained in terms of the lattice parameters as outlined in the framework above for the right eigenvector. Note that the components of the left and right eigenvectors are not the same in general. Moreover, the typical density at site $i$ associated to a given current fluctuation can now be explicitly calculated using the definition $\rho_i(\lambda)=\langle\psi_i| n_i |\psi_i \rangle$, where $n_i$ is the diagonal operator for the number of particles on site $i$. This leads to a relation between densities and fugacities similar to the grand canonical identity~(\ref{grandcanrel-fugdens}) for the steady state, but with the replacement of $z_i$ by $\ti{z}_i\hat{z}_i$. In particular, for the interaction $w_n=n$ the site density is determined by $\rho_{i}(\lambda)=\hat{z}_{i}\ti{z}_{i}$ which reduces in the stationary state to $\rho_{i}(0)=z_i$ since $\ti{z}_i$ equals unity for $\lambda=0$. For bounded $w_n$, the product $\langle \psi|\psi \rangle$ might diverge which again generically indicates condensation~\cite{harris2005}.

In the following we show how to extend the formalism presented here to study the ZRP on different 2-$d$ lattice geometries. In particular, we need to study the influence of the underlying lattice structure on current fluctuations in large finite lattices. To do this, we will modify the Hamiltonians with factors $\exp(\mp\lambda_k)$ to count simultaneously the number of jumps along the corresponding positive or negative $k$-directions.

Specifically, the modified Hamiltonians for the systems shown in Fig.~\ref{lattices} are given explicitly by Eq.~(\ref{Hs-modified}) (with $\ti{\lambda}_y=0$) for the square lattice, and by Eq.~(\ref{Ht-modified}) for the triangular lattice. In this manner, for fixed lattice sizes, we will compute the SCGF
\begin{equation}
e_S(\mb{\lambda})=\lim\limits_{t\ra\infty}-\frac{1}{t}\ln \langle \exp\left( -t \mb{\lambda}\cdot\mb{I} \right) \rangle, \label{scgf-sqr}
\end{equation}
for the square geometry and 
\begin{equation}
\begin{aligned}
e_T(\lambda_1,\lambda_2,\lambda_3) =& \lim\limits_{t\ra\infty}-\frac{1}{t}\ln \langle \exp\left( -t (\lambda_1,\lambda_2,\lambda_3) \right.\\
&\cdot \left.(I_1,I_2,I_3) \right) \rangle,
\end{aligned} \label{scgf-tri}
\end{equation}
for the triangular geometry. To avoid confusion, from now on we will use bold characters to denote vectors in Cartesian coordinates, and we will write explicitly the components of the variables in the triangular lattice.

\section[Effect of lattice geometry]{Effect of lattice geometry on current fluctuations} \label{s:sqrtrilatt}

It is known that in some cases the underlying geometry in lattice-based models can have a significant effect on the results obtained from a microscopic point of view. For example in a different context, at the level of the universality of phase transitions in equilibrium systems, some differences were found numerically between square and triangular lattices~\cite{runnels1966} and later predicted theoretically~\cite{baxter1980}. Here, we use the quantum Hamiltonian formalism to calculate the RF of the global current in the two geometries shown in Fig.~\ref{lattices} and thus investigate the influence of the lattice on the dynamical properties of the ZRP. 

One reason we are interested in the RF of particle current fluctuations and their associated ODP is to confirm that the expressions from both lattices converge to the same function and recover the hydrodynamic result under the appropriate scaling. We shall give more details of the scaling in subsection~\ref{ss:hydolimandodp}, but first we calculate the SCGF for finite lattices using the microscopic approach introduced above.

\subsection{General solution on square and triangular lattices}

As mentioned above, to calculate the probability of fluctuations of the global current, we have to measure the number of jumps throughout the lattice in the time interval $[0,t]$. The modified eigenvector, $|\psi\rangle$, associated to the lowest eigenvalue of the modified Hamiltonian, $\hat{H}$, obeys relations analogous to~(\ref{upright}) and~(\ref{downright}) with the modified fugacities $\hat{z}_{j,i}$.

For each lattice we apply the corresponding $\hat{H}$, (\ref{Hs-modified}) or~(\ref{Ht-modified}), to $|\psi\rangle$ and use the eigenvector condition to obtain that the coefficients of the matrices $d_{j,i}$ in the resulting expression have to vanish. This leads to the recursion relation for the modified fugacities of the right eigenvector
\begin{equation}
Q \hat{z}_{j,i+1} + (Y-R) \hat{z}_{j,i} + P \hat{z}_{j,i-1} = 0 \label{differenceeq}
\end{equation}
with boundary conditions
\begin{equation}
\begin{aligned}
Q \hat{z}_{j,2} + \left(Y-R_{l}\right) \hat{z}_{j,1} + A_l&=0\\
A_r +\left(Y-R_{r}\right) \hat{z}_{j,L} + P \hat{z}_{j,L-1}&=0,
\end{aligned} \label{left-rightbcs}
\end{equation}
for the left- and right-hand side respectively. Here, the uppercase parameters for the triangular lattice correspond to the effective bulk rates
\begin{eqnarray}
P &=& p_2 e^{-\lambda_2} + p_3 e^{-\lambda_3}\\
Q &=& q_2 e^{\lambda_2} + q_3 e^{\lambda_3}\\
Y &=& p_1 e^{-\lambda_1} + q_1 e^{\lambda_1},
\end{eqnarray}
boundary injection rates 
\begin{eqnarray}
A_l &=& \sum_{k=1}^{2}\alpha_k e^{-\lambda_k}\\
A_r &=& \sum_{k=1}^{2} \delta_k e^{\lambda_k},
\end{eqnarray}
and site exit rates 
\begin{eqnarray}
R_l &=& \sum_{k=1}^{3}p_k + q_1 + \gamma_2 + \gamma_3\\
R_r &=& \sum_{k=1}^{3}q_k + p_1 + \beta_2 + \beta_3 \\
R &=& \sum_{k=1}^{3}\left(p_k + q_k\right).
\end{eqnarray}

It can be checked that the same difference equation and boundary conditions are obtained for the square lattice with coefficients: 
\begin{equation}
\begin{aligned}
P &= p_x e^{-\lambda_x}\\ 
Q &= q_x e^{\lambda_x}\\
R &= p_x + q_x + p_y + q_y\\
R_l &= p_x + \gamma + p_y + q_y
\end{aligned}
\quad
\begin{aligned}
R_r &= \beta + q_x + p_y + q_y\\
Y &= p_y e^{-\lambda_y} + q_y e^{\lambda_y}\\
A_l &= \alpha e^{-\lambda_x}\\
A_r &= \delta e^{\lambda_x}.
\end{aligned}
\end{equation}
Here we have omitted the subindices of the boundary parameters as particle jumps in and out of the system contribute only to the current in the $x$-direction.

We make use of the periodic boundary conditions to argue that we only have to solve Eq.~(\ref{differenceeq}) and~(\ref{left-rightbcs}) for a single row. In other words, since the fugacities are invariant in the $y$-direction the system can be treated as a quasi-one-dimensional chain. The difference equation~(\ref{differenceeq}) can be solved exactly, but the expressions are too cumbersome to handle; in practice we use a computer algebra package to calculate exact numerical values for the modified fugacities. An analogous calculation is required to find the components of the left eigenvector, $\langle\psi|$, using relations~(\ref{upleft}) and~(\ref{downleft}). As we shall demonstrate, the seemingly technical analysis of this eigenproblem allows us to investigate both the probability of given current fluctuations (since the eigenvalue generically gives the SCGF) and the mechanisms leading to them (since the modified fugacities can be related to densities).

The results of the following subsections are based on the fact that the ground state of the modified Hamiltonians~(\ref{Hs-modified}) and~(\ref{Ht-modified}) for these lattices can straightforwardly be written in terms of the modified fugacities such that the SCGF for the triangular lattice is given by
\begin{equation}
e_T(\lambda_1,\lambda_2,\lambda_3)=L \sum\limits_{k=2}^{3} \left( \alpha_k +\delta_k -\gamma_k e^{\lambda_k} \hat{z}_1 - \beta_k  e^{-\lambda_k}\hat{z}_L \right), \label{scgf-example}
\end{equation}
whereas for the SCGF of the square lattice we have
\begin{equation}
e_S(\mb{\lambda})=L \left( \alpha +\delta -\gamma e^{\lambda_x} \hat{z}_1 - \beta  e^{-\lambda_x}\hat{z}_L \right). \label{scgf-example1}
\end{equation}
Note that due to the symmetry imposed by the periodic boundary conditions in the \emph{y}-direction only the second subindex, related to the $x$-direction, is needed to identify the fugacities. Finally, as mentioned above, we can calculate the average $\langle \psi | n_i |\psi \rangle$ with the modified eigenvectors, which corresponds to the ODP from a microscopic point of view.

\subsection{Matching diffusive processes in square and triangular lattices}

Before we can compare the solutions obtained from the square and triangular lattices, it is necessary to choose carefully the bulk and boundary hopping rates in order to achieve an equivalent behaviour in both systems.

Firstly, since we are interested in modelling diffusive dynamics in the hydrodynamic limit, we consider symmetric hopping rates $q_k=p_k$. Additionally, we match the extraction boundary rates to the bulk hopping rates $\gamma_k=\beta_k=p_k$ so that the boundaries act simply as reservoirs. 

Now to obtain a mapping for the bulk hopping rates between the triangular and square lattices, we need to equate the particle transport bearing in mind the lattice spacing. In other words for our choice of diffusive dynamics, we have to equate the mean square displacement in both lattices. Mathematically, this implies that the hopping rates of the triangular lattice are mapped to the square lattice via the relations
\begin{equation}
\begin{aligned}
p_x&=p_2\cos^2\phi+p_3\cos^2\phi \\
p_y&=p_1+p_2\sin^2\phi+p_3\sin^2\phi,
\end{aligned}\label{rtsrel}
\end{equation}
where $\phi=\pi/6$. 

To obtain diagonal matrices for the diffusion and mobility coefficients as required for the process on the square lattice, we need to identify $p_3=p_2$. Such a choice of hopping rates leads to the simplified mapping
\begin{equation}
\begin{split}
p_1&=p_y-\frac{p_x}{3}\\
p_2&=\frac{2 p_x}{3}\\
\end{split}
\quad\text{or}\quad
\begin{split}
p_x&=\frac{3p_2}{2}\\
p_y&=p_1+\frac{p_2}{2}.
\end{split}\label{hoppingrels}
\end{equation}
Now we can check that for an isotropic choice of hopping rates in the square lattice (i.e., $p_y=p_x$), our transformations yield isotropic rates in the triangular geometry. Similarly, we can use~(\ref{hoppingrels}) to confirm $2(p_x+p_y)=2(p_1+p_2+p_3)$, so the exit rate from a bulk site is the same in both lattices.

The same reasoning can be used to determine the mapping of the boundary rates, which yields the analogous expressions 
\begin{equation}
\begin{split}
\alpha_k &= \frac{2\alpha}{3}
\end{split}
\qquad
\begin{split}
\delta_k &= \frac{2\delta}{3}
\end{split}\label{hoppingrelsbc}
\end{equation} 
These relations conserve the injection-extraction ratios, $\alpha/p_x = \alpha_2/p_2 = \alpha_3/p_3$ (and analogously for the RHS boundary), which is equivalent to reservoirs with the same fugacity, $z_l=\alpha_k/p_k$ and $z_r=\delta_k/p_k$, for both lattices.

Using the transformation relations~(\ref{hoppingrels}) and~(\ref{hoppingrelsbc}), particle diffusion on the two lattices can be related. To convert current fluctuations in the triangular lattice to Cartesian coordinates, we have to specify how to count particle jumps with the quantum Hamiltonian formalism. An appropriate relation between the triangular and square geometries can be obtained by noticing that currents in the triangular lattice have components
\begin{equation}
\begin{aligned}
j_x&=j_2\cos\phi+j_3\cos\phi\\
j_y&=j_1+j_2\sin\phi-j_3\sin\phi.
\end{aligned}\label{currsqrtri}
\end{equation}
Then, we can use the chain rule on Eq.~(\ref{eq8}) together with the relations~(\ref{currsqrtri}) to obtain
\begin{equation}
\begin{aligned}
\lambda_1 &=\lambda_y\\
\lambda_2 &=\lambda_x \cos\phi + \lambda_y \sin\phi\\
\lambda_3 &=\lambda_x \cos\phi - \lambda_y \sin\phi,
\end{aligned}\label{lambdamapping}
\end{equation}
which are the appropriate conjugate variables to compare the number of particle jumps on a triangular lattice with those in a square lattice.

\subsection{Hydrodynamic limit and optimal density profiles} \label{ss:hydolimandodp}

In this subsection, we focus on explaining the scaling of our results obtained from the microscopic approach with the goal of determining the influence of the underlying lattice geometry for large systems. We will also compare the microscopic results with those obtained in Appendix~\ref{a:mft-calculations} using the MFT. We begin by discussing the scaling for the SCGF~(\ref{scgf-sqr}) of the square lattice, as it is more intuitive than the triangular geometry which will be explained immediately after. Later, we will explain how to scale the density profiles, which is done in a similar way.

As hinted in the introduction, to obtain the hydrodynamic limit of diffusive systems starting from a microscopic approach, a spatial and temporal rescaling is needed. Specifically, space is scaled as $1/L$ and time as $1/L^2$, where $L$ is the linear size of the microscopic system (see e.g.~\cite{derrida2007}). In the quantum Hamiltonian formalism, this leads to dividing the conjugate parameters by the number of bonds in the corresponding direction, as we measure particle jumps throughout the lattice. In $d$ dimensions, the temporal and spatial rescaling also combine to give a factor of $L^{2-d}$ multiplying the SCGF~\cite{hurtado2014} but this reduces to unity for our square 2-$d$ system. Hence we expect the macroscopic SCGF, $e(\mb{\lambda})$, to be given by the limit
\begin{equation}
e\left(\mb{\lambda}\right) = \lim\limits_{L\ra\infty} \frac{L+1}{L} \, e_S\left( \frac{\lambda_x}{L+1} , \frac{\lambda_y}{L} \right). \label{scgfscaling-squarelattice}
\end{equation}
Here we have included an additional factor of $(L+1)/L$ to remove finite size effects in the $x$-direction of small lattices; the large-$L$ limit is clearly unaffected by this.

To obtain the correct scaling for the triangular lattice, we have to remember that the length of the lattice in the $x$-direction is smaller than in the square lattice, being multiplied by a factor of $\cos\phi$. This can be compensated by a modified-length spatial and temporal scaling, leading us to the limit
\begin{equation}
\begin{aligned}
e\left(\mb{\lambda}\right) &= \lim\limits_{L\ra\infty} \frac{L+1}{L\cos^2\phi} \, e_T\left( \frac{\lambda_1}{L\cos\phi} , \right. \\
 & \qquad  \left. \frac{\lambda_2}{(L+1)\cos\phi} , \frac{\lambda_3}{(L+1)\cos\phi} \right),
\end{aligned}\label{scgfscaling-trilattice}
\end{equation}
where $\lambda_1$, $\lambda_2$ and $\lambda_3$ are given by~(\ref{lambdamapping}). We have checked that the additional $\cos\phi$ factors in the argument of $e_T$ can be removed by considering a lattice with $L/\cos\phi$ sites in the $x$-direction (i.e., spatial length $L$) and $L$ sites in the $y$-direction. However, this produces more complicated finite size effects since the number of sites has to be rounded to an integer.

In the present work, we show the results for the interaction $w_n=n$, but we have checked also the case with $w_n=w$ ($w$ constant) within the fluid regime which leads to similar findings. Indeed, the SCGF is invariant with respect to the interaction as long as there is no condensation~\cite{chernyak2011}, but the relation between the density and the fugacity (and hence between boundary rates and reservoir densities) does change. The special case of $w_n = n$ is particularly illuminating because densities then turn out to be proportional to fugacities, so calculations of the latter offer direct physical insight into the optimal profiles. In addition to verifying the hydrodynamic limit with the above scaling, studying the SCGF also provides a convenient way to test the AFR from a microscopic point of view.

\begin{figure}[t]
	\begin{center}
		\includegraphics[width=3.4in]{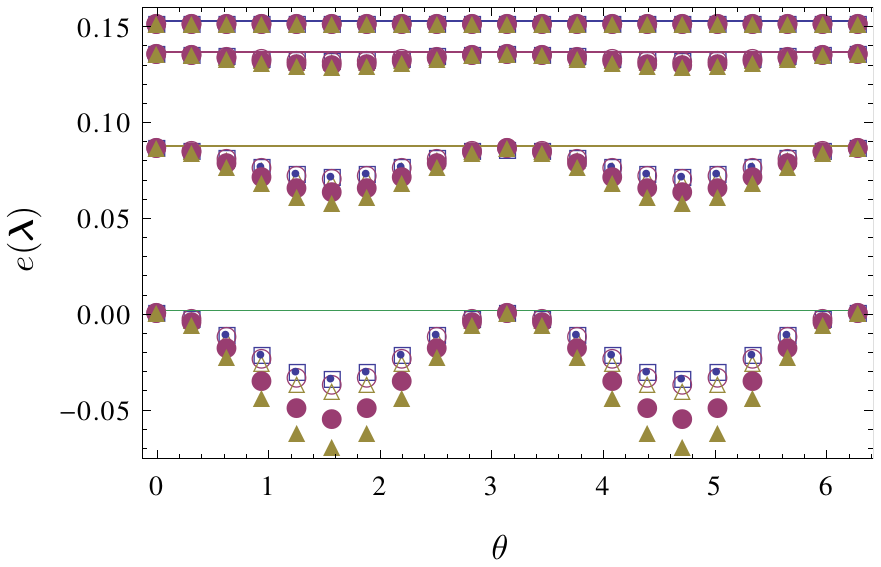}
	\end{center}
	\caption[AFR for the SCGF of square and triangular lattices, and macroscopic approach]{(Color online) SCGF for $\mb{\lambda}(\theta)$ on concentric ellipses around $\mb{E}=1/2\left(\ln(\alpha\beta/(\gamma\delta)),0\right)$ with principal axes in the $x$-direction of length $r_x=\{0,0.266,0.533,0.8\}$ from top to bottom. Square lattice with $L=\{6\,(\blacktriangle),10\,(\CIRCLE),10^5\,(\bullet)\}$, triangular lattice with $L=\{6\,(\bigtriangleup),10\,(\ocircle),10^5\,(\square)\}$, and macroscopic approach (solid line). Hopping rates given by Eq.~(\ref{model-parameters}).}
	\label{scgfafrsqrtrimacro}
\end{figure}

In Fig.~\ref{scgfafrsqrtrimacro} we plot the RHS of~(\ref{scgfscaling-squarelattice}) and~(\ref{scgfscaling-trilattice}) for increasing lattice sizes and values of $\mb{\lambda}$ for which Eq.~(\ref{ellipses-lambda}) is satisfied. We assume bulk and boundary hopping rates, 
\begin{equation}
\begin{split}
\alpha &=1/2\\
\gamma &=\beta=1\\
\delta &= 1/10
\end{split}
\qquad
\begin{split}
p_x &=q_x=1\\
p_y &=q_y=1/2,
\end{split}\label{model-parameters}
\end{equation}
which make particle diffusion anisotropic; of course, it is also possible to test the IFR for isotropic rates. The rescaled microscopic results are compared with the numerical Legendre transform of the macroscopic RF obtained in Appendix~\ref{a:mft-calculations} with reservoir densities $\rho_l=\alpha$ and $\rho_r=\delta$. We can see that both microscopic SCGFs converge to the same function when $L\ra\infty$. As might be expected (due to a larger number of bonds), with the triangular lattice the SCGF has a quicker convergence towards the hydrodynamic limit than with the square lattice. However, it can also be observed that this limit does not agree with the result obtained using the MFT under the assumption of \emph{homogeneous} OCPs.

Turning our attention to the AFR, note that in Fig.~\ref{scgfafrsqrtrimacro} we parametrize in polar coordinates (with angle $\theta$) the values $\mb{\lambda}$ for ellipses centred at
\begin{equation}
\mb{E}=\frac{1}{2}\left(\ln\left(\frac{\alpha\beta}{\gamma\delta}\right),0\right) \label{field-sqrlat}
\end{equation}
on which the SCGF is predicted to be constant. As shown in~\cite{villavicencio2014}, here the AFR is satisfied by the macroscopic results for all points on the ellipse but only for certain angles from the microscopic point of view. Importantly, agreement in the hydrodynamic limit of the microscopic SCGFs of the two lattices, indicates that the discrepancies are not related to the underlying structure. We continue to investigate further the ODP from both approaches.

The microscopic ODP is obtained as explained in Section~\ref{s:micapp}, but also have to be rescaled before we can compare them with the macroscopic ODP computed in Appendix~\ref{a:mft-calculations}. Firstly, notice that within the modified Hamiltonian dynamics the ODP at site $i$ is given by the mean occupation, $\langle n_i\rangle$, and has to be compared with the macroscopic ODP at $x=i/L$. Secondly, scaling of the conjugate variables is done the same as in the arguments of~(\ref{scgfscaling-squarelattice}) and~(\ref{scgfscaling-trilattice}). This way, we are able to compare $\rho_S\left(\lambda_x/(L+1),\lambda_y/L\right)$ and $\rho_T\left(\lambda_1/(L\cos\phi),\lambda_2/((L+1)\cos\phi),\lambda_3/((L+1)\cos\phi)\right)$ for the square and triangular lattices, as well as $\rho\left(\mb{J}(\mb{\lambda})\right)$ for the macroscopic approach.

\begin{figure}[t]
	\centering
	\subfigure[][]{\includegraphics[width=3.4in]{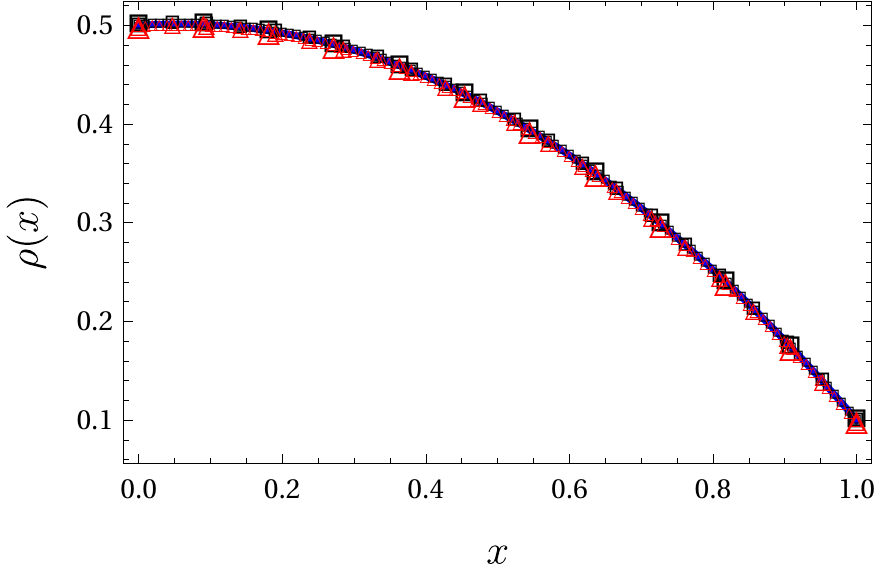}	\label{SqrTriHydDens-top}}
		\quad
	\subfigure[][]{\includegraphics[width=3.4in]{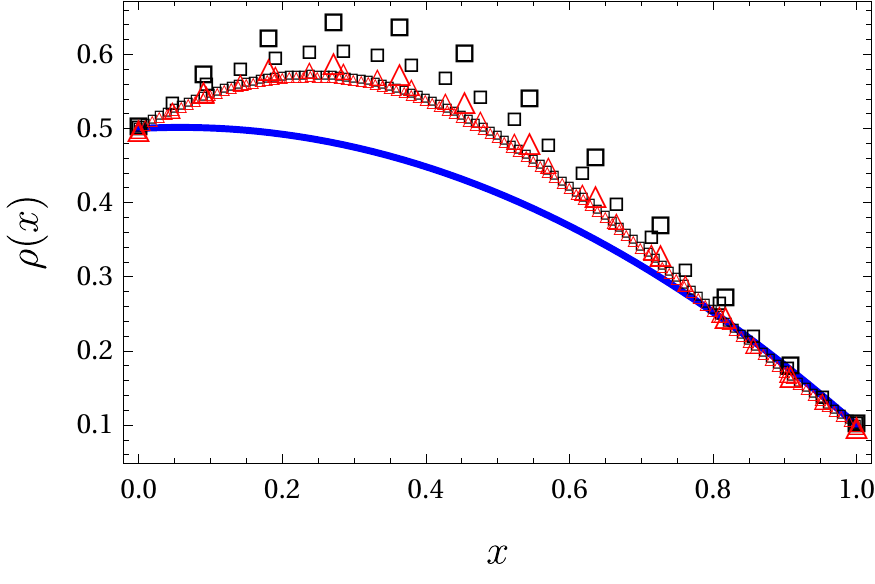}	\label{SqrTriHydDens-bottom}}
	\caption[Optimal density profile: Square lattice, triangular lattice, and MFT prediction]{(Color online) Optimal density profiles on $L\times L$-site lattices. Hopping rates: $\alpha=1/2$, $\delta=1/10$, $p_x=1$, $p_y=1/2$, and $w_n=n$. ODP from microscopic approach on square lattice ($\square$), and triangular lattice ($\triangle$) with symbols of decreasing size for $L=\{10,20,10^5\}$; ODP from macroscopic theory with blue solid line. \subref{SqrTriHydDens-top}~Current fluctuation in $x$-direction: $\mb{\lambda}\simeq(-0.6953,0)$, i.e., $\mb{J}\simeq(0.9523,0)$. \subref{SqrTriHydDens-bottom}~Current fluctuation in $y$-direction: $\mb{\lambda}\simeq(0.8047,-2.1213)$, i.e., $\mb{J}\simeq(0,0.8928)$.}
	\label{SqrTriHydDens}
\end{figure}
Since the ODP has no dependence in the $y$-direction (due to periodic boundary conditions), in Fig.~\ref{SqrTriHydDens} we plot the $x$-projection of the profiles for two values of $\mb{\lambda}$ which are predicted to satisfy the AFR. We have used the same bulk and boundary parameters as for the SCGFs plotted in Fig.~\ref{scgfafrsqrtrimacro}. It can be seen from the upper panel of Fig.~\ref{SqrTriHydDens} that our calculations for both lattices match closely the macroscopic solution (solid line) for $\mb{\lambda}$ in the $x$-direction even for lattices with $L=10$. In the lower panel we choose a current in the $y$-direction only, specifically a current of appropriate magnitude such that the macroscopic ODP remains invariant with respect to the upper panel. The microscopic-approach ODPs from the two lattices converge towards the same function in the hydrodynamic limit, but not to the MFT prediction. This again indicates that the AFR is not exact between these current fluctuations. We suggest that the assumption of a space-homogeneous OCP in the MFT lies behind this discrepancy, and we will investigate it further in the following section by looking for a more detailed structure of the current fluctuations in $L\times L$ square lattices.

\section{Structure of optimal current profiles} \label{s:inhomogcurr}

In this section, we extend our study of global current fluctuations, to gain a fine-grained understanding of the underlying local structure. Specifically, we seek information about the OCP giving rise to a particular global current fluctuation. In contrast to hypothesis \emph{iii}) of the AFR (see Appendix~\ref{a:alternativeafr}), we anticipate finding some spatial dependence (with a similar structure expected for isotropic systems as relevant for the IFR). This conjecture can be motivated by remembering the definition of the global current and the implications of measuring a rare realization of it. When we calculate the RF of a certain fluctuation $\mb{J}$, what we are considering is a space- and time-average of the number of particle jumps throughout the lattice. However, there could be many local current profiles, with different spatial dependence, leading to this average. From all such profiles we want to find the OCP.

\begin{figure}[t]
	\begin{center}
		\includegraphics[width=3.4in]{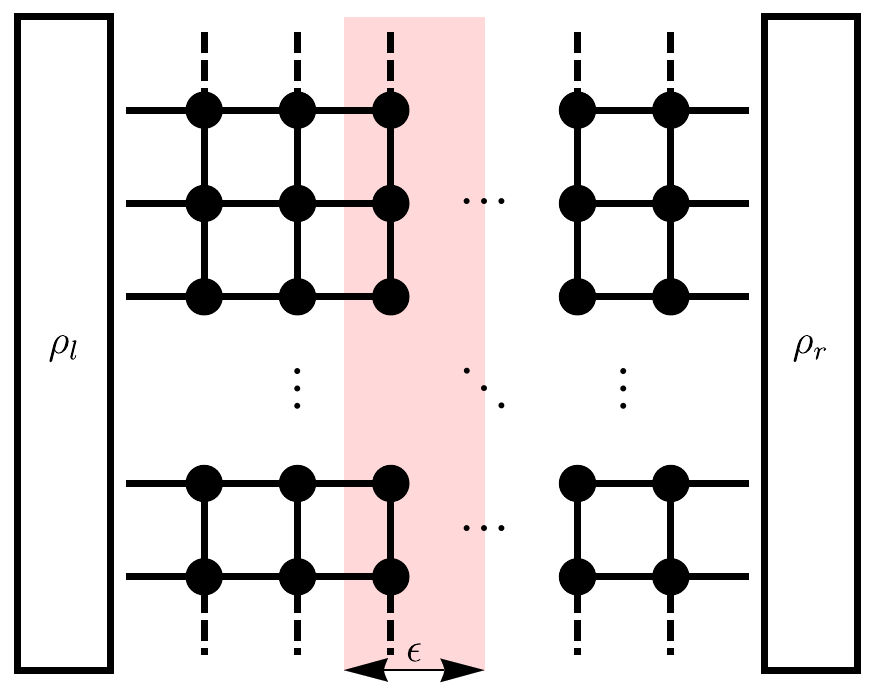}
	\end{center}
	\caption[Local current fluctuations on square lattice]{(Color online) We count particle jumps in the $y$-direction within the highlighted region of width $\epsilon$ and calculate the joint SCGF $e(\ti{\lambda}_y,\mb{\lambda})$.} 
	\label{LxLinhomog}
\end{figure}

In order to gain a deeper understanding of the fluctuations in the 2-$d$ ZRP on a square lattice, we consider the joint probability distribution function (PDF) of a global current and a \emph{local} current in the $y$-direction of a vertical strip, $\mathcal{V}$, as indicated in Fig.~\ref{LxLinhomog}. The relative area of $\mathcal{V}$ is kept constant for all lattice sizes. This implies that the width of $\mathcal{V}$ is made proportional to the lattice length as we increase the number of sites in the system. Thus we anticipate that properly rescaled microscopic results will approach a consistent hydrodynamic limit for increasing $L$. For the purposes of discussing this limit we use the macroscopic notation, but the underlying calculations are still done using the microscopic approach.

To look for the structure of the OCP, we compute the joint RF of the local and global currents,
\begin{equation}
\hat{e}(\ti{J}_y(x_0),\mb{J})=\lim\limits_{tL^d\ra\infty} -\frac{1}{tL^d}\ln\left( P(\ti{J}_y(x_0),\mb{J},t) \right). \label{jointpdf}
\end{equation}
Here, $\mb{J}$ is defined as in the MFT according to Eq.~(\ref{global-current}) and the local current corresponds to 
\begin{equation}
\ti{J}_y(x_0)=\frac{1}{t}\int_{0}^{t}\mathrm{d}\tau\int_{\mathcal{V}}\mathrm{d}\mb{r}j_y(\mb{r},\tau) \label{local-current}
\end{equation}
where $x_0$ is the left boundary of $\mathcal{V}$. We keep a fixed value of $\mb{J}$ and move the location of $\mathcal{V}$ along the \emph{x}-direction in order to capture the statistical behaviour of the local current $\ti{J}_y(x_0)$ in a more detailed way.

As before we first calculate the microscopic SCGF as the lowest eigenvalue of a modified Hamiltonian; here counting the number of particle jumps in $\mathcal{V}$ along the $y$-direction requires us to modify some terms of the stochastic generator with the additional variable $\ti{\lambda}_y$, where the dependence of this modification on $x_0$ is left implicit throughout. Specifically, calculating the modified fugacities involves solving a recursion relation similar to~(\ref{differenceeq}) and~(\ref{left-rightbcs}), where introducing $\ti{\lambda}_y$ imposes the new relation
\begin{equation}
Q \hat{z}_{j,i+1} + (\ti{Y}-R) \hat{z}_{j,i} + P \hat{z}_{j,i-1} = 0, \label{newdifferenceeq}
\end{equation}
for sites within $\mathcal{V}$. Here $\ti{Y}=p_ye^{-\lambda_y-\ti{\lambda}_y}+q_ye^{\lambda_y+\ti{\lambda}_y}$. Taking into account this modification, we were not able to find an exact analytical expression for the SCGF, but we still obtain a complete system of linear equations which can be solved numerically. Finally, the SCGF is rescaled similarly to~(\ref{scgfscaling-squarelattice}) to obtain the hydrodynamic limit.

To probe the spatial dependence of the current profile, we begin by considering how the RF, $\hat{e}_{\mb{J}}\left(\ti{J}_y(x_0)\right):=\hat{e}\left(\ti{J}_y(x_0),\mb{J}\right)$, with \emph{fixed} global current $\mb{J}=(J_x,0)$, changes as $\mathcal{V}$ sweeps the lattice. This is equivalent to calculating the RF of the conditional probability of measuring a local current, given a fixed global current weighted by the probability of that global current, (i.e. $P(\ti{J}_y(x_0)|\mb{J})P(\mb{J})$); the RFs of the conditional and the joint probabilities differ only by a term independent of $\ti{J}_y(x_0)$. In practice, we will focus on studying the corresponding SCGFs.

\begin{figure}[t]
	\centering
	\subfigure[][]{\includegraphics[width=3.4in]{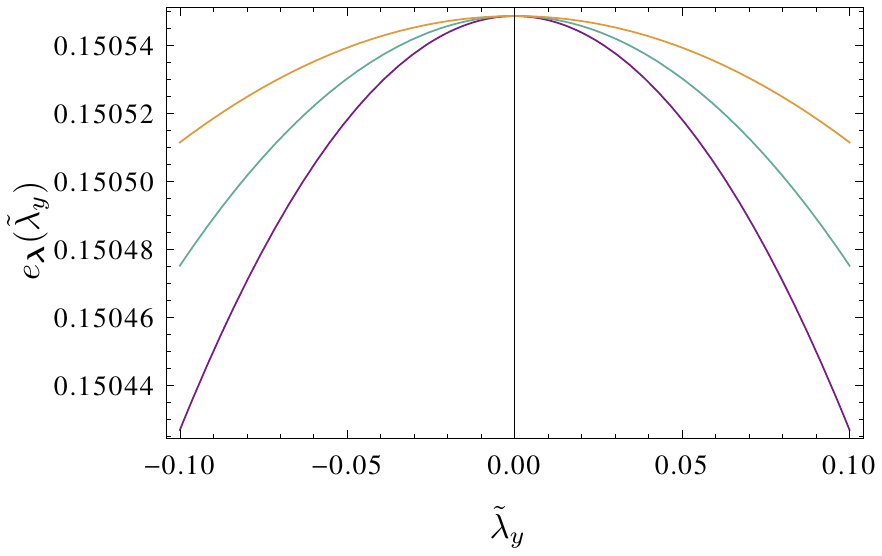}	\label{scgf_slit-top}}
		\quad
	\subfigure[][]{\includegraphics[width=3.4in]{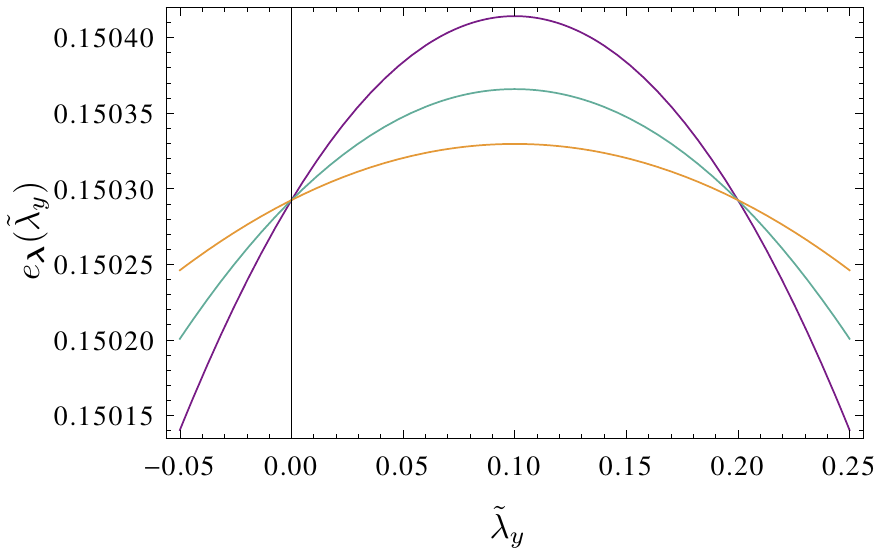}	\label{scgf_slit-bottom}}
	\caption[Current fluctuations in $y$-direction]{(Color online) $e_{\mb{\lambda}}(\ti{\lambda}_y)$ for $\mb{\lambda}$'s predicted to satisfy the AFR, and local conjugate parameter, $\ti{\lambda}_y$, for a slit of relative width $\epsilon=1/20$ with its left boundary located at $x_0=\{0,2/5,4/5\}$ (parabolas of increasing broadness); rescaled microscopic results from a lattice with $L=10^3$. Same boundary rates, bulk hopping rates, and $w_n$ as Fig.~\ref{SqrTriHydDens}. \subref{scgf_slit-top}~Current fluctuation in $x$-direction: $r_x=0.1$ and $\theta=\pi$, i.e., $\mb{\lambda}\simeq(0.7047, 0)$. \subref{scgf_slit-bottom}~Current fluctuation in diagonal direction: $r_x=0.1$ and $\theta=5\pi/4$, i.e., $\mb{\lambda}\simeq(0.7340, -0.1)$.}
	\label{scgf_slit}
\end{figure}

In Fig.~\ref{scgf_slit} we plot the joint SCGF of $\ti{\lambda}_y$ and fixed $\mb{\lambda}$, $e_{\mb{\lambda}}(\ti{\lambda}_y)$. From the top panel (where $\mb{\lambda}$ is fixed in the $x$-direction) we can see that at the three chosen positions on the lattice, we have $\partial e_{\mb{\lambda}}(\ti{\lambda}_y)/\partial \ti{\lambda}_y \vert_{\ti{\lambda}_y=0}=0$, implying that the local mean current in the $y$-direction vanishes in all cases. Additionally, the SCGF becomes broader as $\mathcal{V}$ approaches the right boundary (broader SCGF means that the absolute value of the second derivative is smaller). Taking into account that the variance of the local current can be calculated as 
$-\partial^2 e_{\mb{\lambda}}(\ti{\lambda}_y) /\partial\ti{\lambda}_y^2\vert_{\ti{\lambda}_y=0}$,
implies that $\ti{J}_y(x_0)$ is less prone to fluctuations near the right reservoir. This can be understood physically as having a higher chance to see variations of the current where the sites have more particles available, as long as the system is in the fluid state. The fact that the variance of $\ti{J}_y(x_0)$ is spatially dependent, means for its conditional PDF that in general when $x_0 \neq x_0'$
\begin{equation}
P(\ti{J}_y(x_0)|\mb{J})\neq P(\ti{J}_y(x_0')|\mb{J}) \label{bdwn1}
\end{equation}
even if $\mb{J}=(J_x,0)$.

In the same manner, we calculate the SCGF assuming a fixed global current fluctuation away from the \emph{x}-axis ($\theta=5\pi/4$). The result is shown in the bottom panel of Fig.~\ref{scgf_slit}. In this case, we recover the same behaviour as before for the broadness of the SCGF, but with the maximum of the SCGF displaced. One can easily check that the displacement corresponds to 
\begin{equation}
\ti{E}_y=-\lambda_y, \label{fieldlocalcurrent}
\end{equation}
which can be seen as an artificial field in the driven dynamics caused by the global conditioning. Significantly, taking the derivative of the SCGF we have
\begin{equation}
\E[\ti{J}_y(x_0)|\mb{J}] = \left.\frac{\partial e_{\mb{\lambda}}(\ti{\lambda}_y)}{\partial \ti{\lambda}_y } \right\vert_{\ti{\lambda}_y=0} \neq 0 \label{local-y-current}
\end{equation}
which is no longer constant at different locations of $\mathcal{V}$, meaning that in general when $x_0\neq x_0'$
\begin{equation}
\E[\ti{J}_y(x_0)|\mb{J}]\neq \E[\ti{J}_y(x_0')|\mb{J}] .\label{bdwn2}
\end{equation}
Here, $\E[\cdot|\cdot]$ denotes the conditional expectation of a local current. The inequality~(\ref{bdwn2}) implies physically that for a specific global current, the average local current at $x_0$ and at $x_0'$ is not the same, causing the OCP of the corresponding $\mb{J}$ to be inhomogeneous. 

\begin{figure}[t]
	\centering
	\subfigure[][]{\includegraphics[width=3.4in]{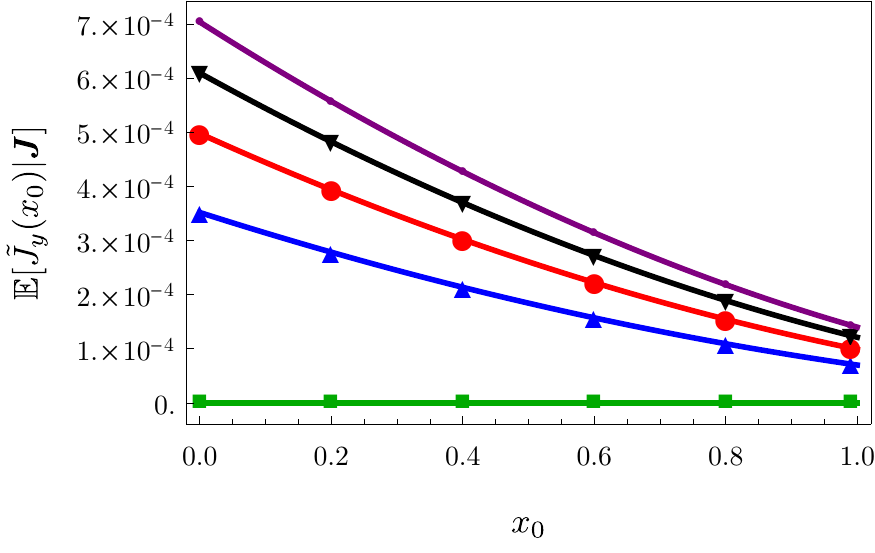}	\label{meanlocalcurrents-top}}
		\quad
	\subfigure[][]{\includegraphics[width=3.4in]{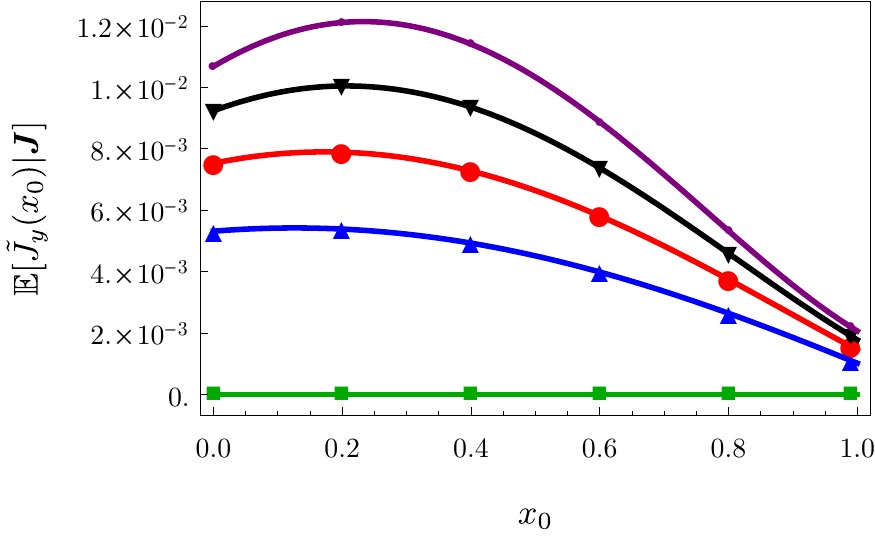}	\label{meanlocalcurrents-bottom}}
	\caption[Local current fluctuations in $y$-direction]{(Color online) Local mean currents in $y$-direction as a function of $x_0$ for fixed $\mb{J}$ predicted to satisfy the AFR. Symbols show numerical values for $\mathcal{V}$ of relative width $\epsilon=1/100$ at $x_0\in\{0,1/5,2/5,3/5,4/5,99/100\}$ rescaled from a lattice with $L=10^3$. Solid lines show interpolation with fourth degree polynomial. Boundary and bulk hopping rates given by~(\ref{model-parameters}). Global currents at angles $\theta=\{0\, (\blacksquare),\pi/6\,(\blacktriangle),\pi/4\,(\CIRCLE),\pi/3\,(\blacktriangledown),\pi/2\,(\bullet) \}$ on ellipses satisfying~(\ref{currents-ellipse}) passing through~\subref{meanlocalcurrents-top} $\mb{J}\simeq (0.0448,0)$ and~\subref{meanlocalcurrents-bottom} $\mb{J}\simeq (0.9522,0)$.}
	\label{meanlocalcurrents}
\end{figure}

To see in more detail the implication of the inequality~(\ref{bdwn2}), we plot in Fig.~\ref{meanlocalcurrents} the local mean current in the $y$-direction for different global current fluctuations predicted to satisfy the AFR~(\ref{afr-rf1}). Namely, we fix $\mb{\lambda}$ on ellipses centred at the constant field~(\ref{field-sqrlat}) and obtain $\E[\ti{J}_y(x_0)|\mb{J}]$ along the lattice. In particular, we have chosen $\mb{\lambda}$ at angles $\theta=\{\pi,7\pi/6,5\pi/4,4\pi/3,3\pi/2 \}$ belonging to ellipses where the distance to the centre in the $x$-direction (i.e., at $\theta=0$) is ${r_x}=\{0.1,1.5\}$. By taking values in the lower left quarter of the ellipse around $\mb{E}$, we obtain currents in the upper right plane after being mapped by the Legendre transform (as reported in Fig.~\ref{meanlocalcurrents}). Here, we can see that the mean current is homogeneous only when a fluctuation of the global current is precisely in the $x$-direction.

Indeed, the inhomogeneity of the local current is consistent with the action functional of the macroscopic fluctuation theory (see Eqs.~(\ref{macro-rf}) and~(\ref{lagrangian})): notice that the RF is inversely dependent on the mobility coefficient, $\sigma(\rho)$. In particular, in the ZRP with interaction $w_n=n$, $\sigma(\rho)\propto\rho$ so the mobility is an increasing function of the density which implies that it is more cost-effective for the system to generate a current fluctuation where it has a higher density, typically near the left reservoir. This argument can also be made for other ZRPs where the mobility coefficient increases with the density, e.g., with interaction $w_n=w$ (constant $w$) the mobility coefficient is related to the density according to $\sigma(\rho)\propto w\rho/(\rho+1)$.

We have also checked that adding the mean value of the local currents on disjoint regions $\mathcal{V}$ (covering the whole lattice), is consistent with the value fixed for the $y$-component of the global current. This indicates that by considering smaller widths the local mean current profile should converge to the OCP. The analysis of this section therefore suggests that the OCP can be space dependent, which could be responsible for the discrepancies between the microscopic and (current-homogeneous) macroscopic approaches, as well as the fact that the AFR (and the IFR) is not exactly satisfied for currents in the $y$-direction.

\section{Discussion and outlook}\label{s:outlook}

We have studied the ZRP on square and triangular lattice geometries calculating exactly the fugacities throughout the lattice, the SCGF for global current fluctuations, and the density profiles associated to such fluctuations. We also used these results to test a recently predicted symmetry for anisotropic systems (the AFR). Since the ZRP we studied is solved analytically, our results have an advantage compared to other studies of the same class of models where numerical simulations are needed to test convergence towards macroscopic predictions. For example, in~\cite{hurtado2011} the IFR (for isotropic systems) was tested using the Kipnis-Marchioro-Presutti (KMP) process and a hard-disk fluid. In particular, despite using an efficient algorithm, the KMP process was simulated for a maximum lattice size of $L=32$ (i.e. $32^2$ sites).

In~\cite{villavicencio2014} we raised as an observation that given the large lattice sizes considered, up to $L=10^5$, the SCGF and ODPs obtained from the microscopic approach did not seem to converge exactly to the macroscopic prediction. A similar result was obtained here for the triangular lattice, with a quicker convergence of the microscopic results to the same hydrodynamic limit as observed with the square geometry. We believe the discrepancy with the macroscopic prediction is caused by the fact that, in the MFT, the OCP was assumed to be spatially homogeneous. Our analysis here of the local structure of the current fluctuations (within the quantum Hamiltonian formalism) indicates that such an invariance does not hold in general. In fact, one could also relax the assumption of space homogeneity in the MFT and it would be interesting to check the resulting OCP, ODP, and SCGF with the hydrodynamic limit of our results. (Significant work in this direction has already appeared since the submission of our original preprint~\cite{perez-espigares2015a}.)

Crucially, the hypothesis of spatial invariance of the OCP is also used in the derivation of the AFR (and the original IFR). The finding of spatial inhomogeneity thus explains the fact, that for fluctuations away from the field direction, the AFR does not hold exactly in this model. However, we still expect some kind of fluctuation relation along the lines of the AFR without assuming homogeneous OCPs. Such a generalization would presumably not have the same simple structure of~(\ref{afr-rf1}) and~(\ref{currents-ellipse}) but relate only local rotations of the current (compare with the discussion for the IFR in~\cite{hurtado2011,hurtado2014}). We emphasize that the usual AFR is still significant for experiments~\cite{kumar2015}, because it is a good approximation for fluctuations close to the forward direction and therefore enables the testing of fluctuation symmetries without the need to measure rare backward fluctuations. Furthermore, for systems with periodic boundary conditions in \emph{every} direction, the OCP is not expected to have any local structure, so this type of spatial fluctuation relation should be exactly satisfied~\cite{perez-espigares2015}.

Finally, we point out that knowing the local structure can give information about the mechanism that generates a global current fluctuation. In general, rather than creating a global current fluctuation by a homogeneous contribution of particle jumps throughout the system, larger local currents are produced where the mobility is larger. For example, in the fluid regime of the ZRP this happens where more particles are available; in contrast, for the simple symmetric exclusion process we would expect to see larger contributions to the global current for intermediate densities. It would be worthwhile to extend this picture to cases where there are dynamical phase transitions~\cite{harris2006,bodineau2005,hurtado2011b} leading to long-term accumulation of particles within the lattice. Further open questions relate to systems with non-diagonal diffusivity and mobility matrices (for the triangular lattice this can be achieved by setting $p_2\neq p_3$), as well as more general anisotropy with different physical processes in each direction. Experimental tests of fluctuation relations in such situations would also be very enriching.

\section*{Acknowledgements}
We thank Hugo Touchette for discussions during the early stages of this work. RJH is also grateful for the hospitality of the National Institute for Theoretical Physics (NITheP) Stellenbosch during part of the manuscript preparation. RVS was supported by the CONACYT Scholarship scheme.

\appendix
\section[Derivation of the AFR]{Derivation of the anisotropic fluctuation relation} \label{a:alternativeafr}

In this section, for completeness, we show a derivation of the AFR~\cite{villavicencio2014} illustrating explicit details for the minimization of the action functional for current fluctuations. This fluctuation relation generalizes the IFR~\cite{hurtado2011} to anisotropic systems, and was recently derived in~\cite{villavicencio2014} under the hypotheses that the system satisfies: \emph{i}) reversibility and local detailed balance, \emph{ii}) time-invariance of the ODP and OCP, and \emph{iii}) space-invariance of the OCP. We take as a framework the MFT to study systems satisfying the continuity equation
\begin{equation}
\frac{\partial \rho(\mb{r},t)}{\partial t}-\nabla\cdot\mb{j(\mb{r},t)}=0, \label{continuity-equation}
\end{equation}
where $\rho(\mb{r},t)$ and $\mb{j}(\mb{r},t)$ are the local particle density and local current respectively. Here we consider for the space variable, the $d$-dimensional unit interval $\Omega=[0,1]^d$ which leads to the LDP for fluctuations of the global current as stated in Eq.~(\ref{ldp}). As in the main text, we consider a diffusive system in contact with two particle reservoirs with densities obeying $\rho_l>\rho_r$ in the $x$-direction, and periodic boundary conditions in every other direction.

According to the MFT, to compute the macroscopic RF we have to minimize~\cite{bertini2005a,bertini2006}
\begin{equation}
\hat{e}(\mb{J})=\min\limits_{\rho,\mb{j}} \frac{1}{t}\int_0^t \mathrm{d}\tau \int_\Omega \mathrm{d}\mb{r} \mathcal{L}\left(\tau,\mb{r},\rho,\nabla\rho\right), \label{macro-rf}
\end{equation}
with Lagrangian
\begin{equation}
\mathcal{L}\left(\tau,\mb{r},\rho,\nabla\rho\right)=\frac{\left(\mb{j}(\mb{r},\tau)+D\nabla\rho\right)^T\Sigma\left(\mb{j}(\mb{r},\tau)+D\nabla\rho\right)}{4}.\label{lagrangian}
\end{equation}
Here, we have that the local current is modelled by a deterministic and a stochastic term. The deterministic part, relates the current to the density via Fick's law with diffusivity $D(\rho)$ given by the diagonal matrix with elements $D_{k}(\rho)=\Delta_{k} g(\rho)$. Furthermore, the stochastic term corresponds to white noise $\mb{\xi}(\mb{r},t)$ with covariance $L^{-d}\sigma(\rho)\delta(\mb{r}'-\mb{r})\delta(t'-t)$. Here the mobility coefficient is given by the diagonal matrix $\sigma(\rho)$ with elements $\sigma_k(\rho)=\Sigma_k(\rho)^{-1}=\Lambda_k^{-1} f(\rho)$. Note that we have assumed the diffusivity and mobility matrices can be factorized as a matrix of constant coefficients times a function of the density. The physical meaning of such a factorization is that particles diffuse at different rates in different directions but obey a single type of process. In particular, if the constant matrices $\Delta$ and $\Lambda^{-1}$ are both the identity, we have the isotropic dynamics for which the original IFR was derived.

Since minimizing~(\ref{macro-rf}) is still a very general problem, we now use hypotheses \emph{ii}) time-invariant ODP and OCP, and \emph{iii}) space-invariant OCP. Thus, the optimization problem is reduced to
\begin{equation}
\hat{e}\left(\mb{J}\right)=\min\limits_\rho \frac{1}{4}\int_\Omega \mathrm{d}\mb{r} \left(\mb{J}+D\nabla\rho\right)^T \Sigma \left(\mb{J}+D\nabla\rho\right). \label{macro-rf-simpl}
\end{equation}
In contrast to~\cite{hurtado2011,villavicencio2014}, we here explicitly solve the Euler-Lagrange equation
\begin{equation}
\frac{\partial\mathcal{L}}{\partial \rho}-\sum\limits_{k=1}^{d} \frac{\partial}{\partial x_k}\left\{\frac{\partial \mathcal{L}}{\partial \rho_{x_k}^{(1)}}\right\}=0, \label{euler-lagrange}
\end{equation}
where we denote the space-variables in $d$-dimensions by $x_k$ and $\rho_{x_k}^{(n)}=\partial^{(n)}\rho/\partial x_k^n$ with $k\in\{1,...,d\}$. Following this procedure one can compute that
\begin{widetext}
\begin{eqnarray}
\frac{\partial \mathcal{L}}{\partial \rho} &=& \sum\limits_{k=1}^{d} \frac{\left(J_k+D_k \rho_{x_k}^{(1)}\right) \rho_{x_k}^{(1)} \partial_\rho D_k}{2 \sigma_k} - \frac{\left(J_k^2+2J_k D_k \rho_{x_k}^{(1)}+D_k^2 \left(\rho_{x_k}^{(1)}\right)^2\right)\partial_\rho \sigma_k}{4\sigma_k^2} \label{euler-eq1}\\
\frac{\partial}{\partial x_k}\left\{\frac{\partial \mathcal{L}}{\partial \rho_{x_k}^{(1)}}\right\} &=& \frac{D_k \left(\rho_{x_k}^{(1)}\right)^2 \partial_\rho D_k}{\sigma_k} + \frac{D_k^2 \rho_{x_k}^{(2)} + J_k \rho_{x_k}^{(1)} \partial_\rho D_k}{2\sigma_k} - \frac{\left(J_k+D_k \rho_{x_k}^{(1)}\right) D_k \rho_{x_k}^{(1)} \partial_\rho \sigma_k}{2\sigma_k^2}. \label{euler-eq2}
\end{eqnarray}
Then, substituting these two expressions in~(\ref{euler-lagrange}), some simplification leads to the differential equation
\begin{equation}
\sum\limits_{k=1}^d -\frac{2D_k \left(\rho_{x_k}^{(1)}\right)^2 \partial_\rho D_k + 2D_k^2 \rho_{x_k}^{(2)}}{4\sigma_k} + \frac{\left( D_k^2 \left(\rho_{x_k}^{(1)}\right)^2 - J_k^2 \right) \partial_\rho \sigma_k}{4 \sigma_k^2} = 0. \label{eq1}
\end{equation}
\end{widetext}
Moreover, notice that periodic boundary conditions imply for the ODP that $\rho_{x_k}^{(1)}=0$ in all directions except for the one with open boundary conditions ($x_1$), and allow us to replace $2D_k^2\rho_{x_k}^{(2)}$ by $D_k^2\partial_\rho (\rho_{x_k}^{(1)})^2$. Indeed, analogously to~\cite{bodineau2004} in one dimension, we can now integrate Eq.~(\ref{eq1}) with respect to the space variable. This yields the non-linear differential equation
\begin{equation}
\sum\limits_{k=1}^{d} \frac{D_k^2 \left(\rho_{x_k}^{(1)}\right)^2}{4\sigma_k} = \sum\limits_{k=1}^d \frac{J_k^2}{4\sigma_k} + C , \label{ellipse-equation-components}
\end{equation}
where $C$ is a constant of integration related to the boundary conditions. This way, to find the ODP that minimizes~(\ref{macro-rf-simpl}) we have to solve~(\ref{ellipse-equation-components}), from which it can already be seen that for global current fluctuations lying on ellipses (constant first term on the RHS), the ODP will remain \emph{invariant}.

As a final step, note that Eq.~(\ref{ellipse-equation-components}) can be written in a more compact way as
\begin{equation}
\left(D\nabla\rho\right)^T \Sigma \left(D\nabla\rho\right) = \mb{J}^T \Sigma \mb{J} + C. \label{ellipse-equation-compact}
\end{equation}
It is easy to see that taking the difference between the RFs~(\ref{macro-rf-simpl}) of two global current fluctuations for which the RHS of Eq.~(\ref{ellipse-equation-compact}) has the same value, results in the relation
\begin{equation}
\hat{e}\left(\mb{J}\right)-\hat{e}\left(\mb{J}'\right) = \frac{1}{2} \int_\Omega \mathrm{d}\mb{r} \left(D\nabla\rho\right)^T\Sigma \left(\mb{J}'-\mb{J}\right), \label{afr1}
\end{equation}
where $\rho$ is now the ODP. Here, corresponding to hypotheses \emph{ii}) and \emph{iii}) the OCP, $\mb{J}$, can be taken out of the integral. Furthermore, from assumption \emph{i}) it follows that the remaining integral in~(\ref{afr1}) is constant and we define
\begin{equation}
\mb{E} = \frac{1}{2} \int_\Omega \mathrm{d}\mb{r} \left(D\nabla\rho\right)^T\Sigma. \label{constant-field}
\end{equation}
This leads to the anisotropic version of Eq.~(\ref{general-fluctuation-relation}), or in terms of the RF,
\begin{equation}
\hat{e}(\mb{J})-\hat{e}(\mb{J}')=\mb{E}\cdot\left(\mb{J}'-\mb{J}\right) \label{afr-rf1}
\end{equation}
for global currents satisfying
\begin{equation}
\mb{J}^T\Lambda\mb{J}=\mb{J}'^T\Lambda\mb{J}'. \label{currents-ellipse}
\end{equation}
Here the density dependence, $f(\rho)$, of the mobility matrix has cancelled out, and as expected this equation reduces to~(\ref{ellipse-equation}) for isotropic systems which satisfy the IFR. Furthermore, this symmetry also implies that the SCGF satisfies 
\begin{equation}
e(\mb{\lambda})=e(\mb{\lambda}') \label{afr-scgf}
\end{equation}
for ellipses centred around the field $\mb{E}$
\begin{equation}
\left(\mb{\lambda}-\mb{E}\right)^T \Lambda^{-1} \left(\mb{\lambda}-\mb{E}\right) = \left(\mb{\lambda}'-\mb{E}\right)^T \Lambda^{-1} \left(\mb{\lambda}'-\mb{E}\right) \label{ellipses-lambda}.
\end{equation}

In order to test this relation explicitly, one can calculate the ODP and the RF of the global current which has been done in 2-$d$ for the KMP process~\cite{perez-espigares2011a} and the ZRP with interacting and non-interacting particles~\cite{villavicencio2014} (see also Appendix~\ref{a:mft-calculations}).

\begin{widetext}
\section[Quantum Hamiltonians]{Quantum Hamiltonians for square and triangular lattices}\label{a:hamiltonians}

In this section we use the ladder operators~(\ref{ladderops}) to write explicitly the Hamiltonians of the ZRP on the square and triangular lattices shown in Fig.~\ref{lattices}. These Hamiltonians are equivalent to the stochastic generators with the corresponding geometry. At the boundaries, we use generic injection and extraction rates as shown in Fig.~\ref{bdryrates}, whereas hopping rates for bulk sites are taken as shown in the insets of Fig.~\ref{lattices}. For the square lattice we have
\begin{eqnarray}
	-H_{S}&=&\sum\limits_{j=1}^{L} \bigg\{ \alpha \left(a_{j,1}^+ -1\right)
	+ \gamma \left(a_{j,1}^- -d_{j,1}\right)
	 +  \delta \left(a_{j,L}^+ -1\right) 
	+ \beta  \left(a_{j,L}^- -d_{j,L}\right)\nonumber\\
	& + & \sum\limits_{i=1}^{L-1} p_x \left(a_{j,i}^-a_{j,i+1}^+ -d_{j,i}\right) 
	+ q_x \left(a_{j,i}^+a_{j,i+1}^- -d_{j,i+1}\right) \nonumber\\
	& + & \sum\limits_{i=1}^{L} p_y \left(a_{j,i}^-a_{j+1,i}^+ -d_{j,i}\right)
	+ q_y \left(a_{j,i}^+a_{j+1,i}^- -d_{j+1,i}\right)\bigg\},  \label{Hs}
\end{eqnarray}
where assuming periodic boundary conditions in the $y$-direction means that we identify $j=L+1$ with $j=1$. Then, to measure current fluctuations, the Hamiltonian is modified by multiplying the terms corresponding to the bonds where we count particle jumps by the factors $e^{\mp \lambda_k}$. Taking this into account, the modified Hamiltonian for the square lattice measuring current fluctuations globally and in the $y$-direction of region $\mathcal{V}$ (see Fig.~\ref{LxLinhomog}) is given by
\begin{eqnarray}
	-\hat H_{S}&=&\sum\limits_{j=1}^{L} \bigg\{ \alpha \left(a_{j,1}^+e^{-\lambda_x}-1\right)
	+ \gamma \left(a_{j,1}^-e^{\lambda_x}-d_{j,1}\right)
	 +  \delta \left(a_{j,L}^+e^{\lambda_x}-1\right) 
	+ \beta  \left(a_{j,L}^-e^{-\lambda_x}-d_{j,L}\right)\nonumber\\
	& + & \sum\limits_{i=1}^{L-1} p_x \left(a_{j,i}^-a_{j,i+1}^+e^{-\lambda_x}-d_{j,i}\right) 
	+ q_x \left(a_{j,i}^+a_{j,i+1}^-e^{\lambda_x}-d_{j,i+1}\right) \nonumber\\
	& + & \sum\limits_{i=1}^{L} p_y \left(a_{j,i}^-a_{j+1,i}^+e^{-(\lambda_y+\ti{\lambda}_y\mathbb{I}(i,\mathcal{V}))}-d_{j,i}\right)
	+ q_y \left(a_{j,i}^+a_{j+1,i}^-e^{\lambda_y+\ti{\lambda}_y\mathbb{I}(i,\mathcal{V})}-d_{j+1,i}\right)\bigg\}.  \label{Hs-modified}
\end{eqnarray}
Here, $\mathbb{I}(i,\mathcal{V})$ is the indicator function for the sites in $\mathcal{V}$.

\begin{figure}[t]
	\centering
	\subfigure[][]{\includegraphics[width=3.5in]{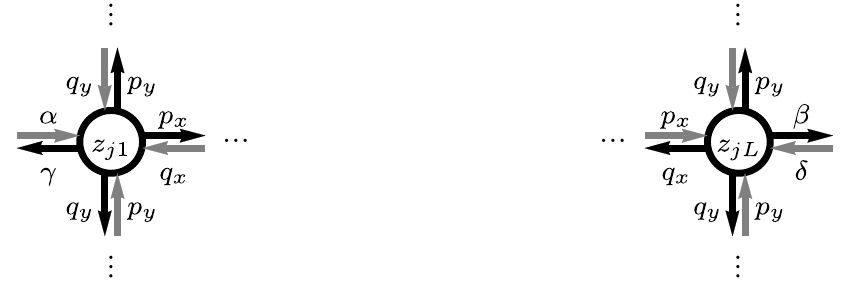} \label{bdryrates-square}}\\ \vspace{.5cm}
	\subfigure[][]{\includegraphics[width=3.5in]{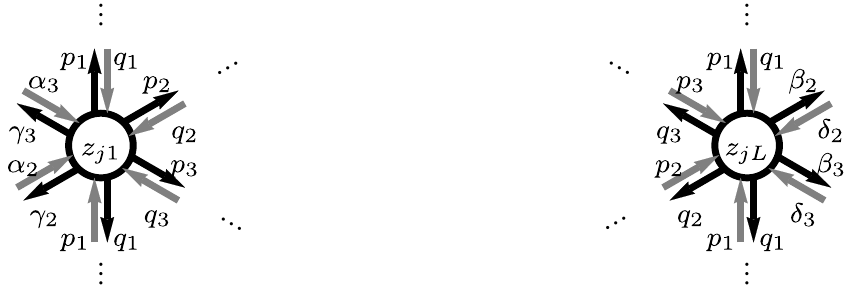} \label{bdryrates-triangular}}
	\caption[Boundary hopping rates for square and triangular lattices]{Hopping rates for boundary sites of~\subref{bdryrates-square} square lattice and~\subref{bdryrates-triangular} triangular lattice. Input rates indicated in gray and output rates in black.}	\label{bdryrates}
\end{figure}
For the triangular geometry the stochastic generator is given by
\begin{eqnarray}
-H_{T} &=& \sum\limits_{j=1}^{L}\left\{\sum\limits_{k=2}^{3} \left[\alpha_k\left(a_{j,1}^+ - 1\right) + \gamma_k\left(a_{j,1}^- - d_{j,1}\right) + \delta_k\left(a_{j,L}^+ - 1\right) + \beta_k\left(a_{j,L}^- - d_{j,L}\right)\right]\right.\nonumber\\
 &+& \sum\limits_{i=1}^{L}p_1\left(a_{j,i}^-a_{j+1,i}^+ - d_{j,i}\right) + q_1\left(a_{j,i}^+a_{j+1,i}^- - d_{j+1,i}\right)\nonumber\\
 &+& \sum\limits_{i=1}^{M-1}\left[p_2\left(a_{j,2i}^-a_{j,2i+1}^+ - d_{j,2i}\right) + q_2\left(a_{j,2i}^+a_{j,2i+1}^- - d_{j,2i+1}\right)\right.\nonumber\\
&\,&\quad + \left.p_3\left(a_{j,2i}^-a_{j+1,2i+1}^+ - d_{j,2i}\right) + q_3\left(a_{j,2i}^+a_{j+1,2i+1}^- - d_{j+1,2i+1}\right)\right]\nonumber\\
&+& \sum\limits_{i=1}^{M}\left[p_2\left(a_{j+1,2i-1}^-a_{j,2i}^+ - d_{j+1,2i-1}\right)+q_2\left(a_{j+1,2i-1}^+a_{j,2i}^- - d_{j,2i}\right)\right.\nonumber\\
&\,&\quad+ \left. p_3\left(a_{j,2i-1}^-a_{j,2i}^+ - d_{j,2i-1}\right) + q_3\left(a_{j,2i-1}^+a_{j,2i}^- - d_{j,2i}\right) \right]\Bigg\}, \label{Ht}
\end{eqnarray}
whereas for the modified Hamiltonian counting particle jumps globally we have
\begin{eqnarray}
-\hat{H}_{T} &=& \sum\limits_{j=1}^{L}\left\{\sum\limits_{k=2}^{3} \left[\alpha_k\left(a_{j,1}^+e^{-\lambda_k}-1\right) + \gamma_k\left(a_{j,1}^-e^{\lambda_k}-d_{j,1}\right) 	
 + \delta_k\left(a_{j,L}^+e^{\lambda_k}-1\right) + \beta_k\left(a_{j,L}^-e^{-\lambda_k}-d_{j,L}\right)\right]\right.\nonumber\\
 &+& \sum\limits_{i=1}^{L}p_1\left(a_{j,i}^-a_{j+1,i}^+e^{-\lambda_1}-d_{j,i}\right) + q_1\left(a_{j,i}^+a_{j+1,i}^-e^{\lambda_1}-d_{j+1,i}\right)\nonumber\\
 &+& \sum\limits_{i=1}^{M-1}\left[p_2\left(a_{j,2i}^-a_{j,2i+1}^+e^{-\lambda_2}-d_{j,2i}\right) + q_2\left(a_{j,2i}^+a_{j,2i+1}^-e^{\lambda_2}-d_{j,2i+1}\right)\right.\nonumber\\
&\,&\quad+ \left.p_3\left(a_{j,2i}^-a_{j+1,2i+1}^+e^{\lambda_3}-d_{j,2i}\right) + q_3\left(a_{j,2i}^+a_{j+1,2i+1}^-e^{\lambda_3}-d_{j+1,2i+1}\right)\right]\nonumber\\
&+& \sum\limits_{i=1}^{M}\left[p_2\left(a_{j+1,2i-1}^-a_{j,2i}^+e^{-\lambda_2}-d_{j+1,2i-1}\right)+q_2\left(a_{j+1,2i-1}^+a_{j,2i}^-e^{\lambda_2}-d_{j,2i}\right)\right.\nonumber\\
&\,&\quad+ \left. p_3\left(a_{j,2i-1}^-a_{j,2i}^+e^{-\lambda_3}-d_{j,2i-1}\right) + q_3\left(a_{j,2i-1}^+a_{j,2i}^-e^{\lambda_3}-d_{j,2i}\right) \right]\Bigg\}. \label{Ht-modified}
\end{eqnarray}
Here we again assume periodic boundary conditions in the $y$-direction, and without loss of generality, an even number of sites $L=2M$.
\end{widetext}

\section{Macroscopic RF and ODP}\label{a:mft-calculations}
In this appendix we show in detail how to calculate, according to the macroscopic fluctuation theory, the ODP and the RF for the 2-$d$ ZRP. We follow~\cite{perez-espigares2011a}, where a similar calculation was done for the 2-$d$ KMP process. Again, we consider open boundary conditions in the $x$-direction and periodic in the $y$-direction. The left- and right-reservoir densities, $\rho_l$ and $\rho_r$ respectively, satisfy the inequality $\rho_l>\rho_r$ which indicates that the NESS has a mean current profile in the rightwards direction. We here assume the same hypotheses used to derive the AFR above (including space-homogeneous OCPs), solving Eq.~(\ref{macro-rf-simpl}) for the RF and Eq.~(\ref{ellipse-equation-components}) for the ODP. 

Firstly, note that the general diffusion and mobility (diagonal) matrices for the ZRP are given by $D(\rho) = \Delta z'(\rho)$ and $\sigma(\rho) = \Lambda^{-1} z(\rho)$ (i.e., $f(\rho)=z(\rho)$ and $g(\rho)=z'(\rho)$) where $z'(\rho)=dz(\rho)/d\rho$ and the components are $\Delta_k=\Lambda_k^{-1}=p_k$~\cite{kipnis1999,bertini2002}. To compute the ODP, we substitute these in Eq.~(\ref{ellipse-equation-components}) which leads to the non-linear partial differential equation
\begin{equation}
\sum\limits_{k=1}^{2} p_{x_k} \frac{z'(\rho)^2}{z(\rho)} \left(\frac{\partial\rho}{\partial x_k}\right)^2 = \sum\limits_{k=1}^{2} \frac{J_{x_k}^2}{p_{x_k} z(\rho)}+4C. \label{optprof0}
\end{equation}
Here, we denoted $x$ by $x_1$ and $y$ by $x_2$. Note that the fugacity $z(\rho)$ and its derivative $z'(\rho)$ take different functional forms according to the type of interaction term $w_n$. Additionally, due to the cylindrical symmetry of the space we can assume that the profiles are flat in the direction of the periodic boundary conditions. Since we assume open boundary conditions in the $x$-direction only, the $k=2$ term on the LHS of Eq.~(\ref{optprof0}) vanishes leading to
\begin{equation}
p_x \frac{z'(\rho)^2}{z(\rho)} \left(\frac{\partial\rho}{\partial x}\right)^2 = \sum\limits_{k=1}^{2} \frac{J_{x_k}^2}{p_{x_k} z(\rho)}+4C. \label{optprof}
\end{equation}
This equation is transformed into a differential equation for the fugacity, which depending on the type of interaction chosen, can be mapped to the density to obtain the optimal profile.

\subsection{Optimal density profile}

In the ZRP, the relation between the diffusivity and mobility matrices allows us to transform Eq.~(\ref{ellipse-equation-components}) into a relation for the fugacity which can be solved independently of the interaction $w_n$. The interaction plays a role when we relate the fugacity to the density, e.g., for non-interacting particles $z(\rho)=\rho$. For now, we eliminate the explicit dependence on the density and write Eq.~(\ref{optprof}) in terms of the fugacity as
\begin{equation}
\left(\frac{\partial z}{\partial x}\right)^2 = \frac{J_{x}^2}{p_x^2}+\frac{J_{y}^2}{p_x p_y}+\frac{4C z}{p_x}. \label{optprof1}
\end{equation}
The non-linearity of this equation requires us to consider two cases. The first one corresponds to a (decreasing) monotonic ODP where the largest density is at the left reservoir, and the second, corresponds to a non-monotonous profile with a maximum fugacity, $z^*$, at a distance $x^*$ from the left particle reservoir. 

Firstly, for the monotonous regime, we have to solve
\begin{equation}
\frac{\partial z}{\partial x} = -\sqrt{a+bz}, \label{optprof-nonint-monot}
\end{equation}
where $a=J_x^2/p_x^2 + J_y^2/p_xp_y$ and $b=4C/p_x$. This leads to the solution
\begin{equation}
z(x)=z_l-x\sqrt{a+b z_l}+\frac{b x^2}{4}, \label{optprof-nonint-monot1}
\end{equation}
where to satisfy the open boundary conditions, $z(0)=z_l$ and $z(1)=z_r$, the constant of integration is determined by
\begin{equation}
b=4\left(z_l+z_r\pm\sqrt{a+4 z_l z_r}\right) \label{intconstant}
\end{equation}
with the negative sign corresponding to the physical solution.

In the non-monotonous regime the optimal profile has a maximum, $z^*=z(x^*)$; the transition to this regime appears when the RHS of~(\ref{optprof-nonint-monot}) vanishes for the first time (i.e., when $x^*=0$ and $z^*=z_l$). Moreover, from~(\ref{optprof-nonint-monot}) we identify $b=-a/ z_l$ and from~(\ref{optprof-nonint-monot1}) we get that the change of regime appears for currents
\begin{equation}
\frac{J_x^2}{p_x}+\frac{J_y^2}{p_y} = 4 z_lp_x\left( z_l- z_r\right). \label{transitioncurr-nonint}
\end{equation}
Clearly, these currents lie on an ellipse and will all have the same ODP.

In the non-monotonous regime, we separate the solution of Eq.~(\ref{optprof1}) into two branches: one to the LHS of $x^*$, and one to the RHS. Due to the non-linearity of this equation the derivative of the profile must be positive when $x<x^*$, and negative when $x>x^*$. Additionally, since $z^*$ is constant for any current on a fixed ellipse, we can use it to replace $b$ by $-a/z^*$. This way, we write the RHS of~(\ref{optprof1}) as $a(1-z/z^*)$ finding that the ODP has fugacities
\begin{equation}
  z(x)= 
  \begin{cases}
    z_l-\displaystyle \frac{ax^2}{4z^*}+x\sqrt{a\left(1-\displaystyle \frac{z_l}{z^*}\right)} & x\leq x^* \\
    z_r-\displaystyle \frac{a(x-1)^2}{4z^*}+(1-x)\sqrt{a\left(1-\displaystyle\frac{z_r}{z^*}\right)} & x> x^*
  \end{cases}.\label{optprof-nonint-nonmonot}
\end{equation}
Here, the maximum fugacity $z^*$ and its position $x^*$ are determined self-consistently resulting in
\begin{eqnarray}
\begin{aligned}
z^*&=\frac{a\left(z_l+z_r+\sqrt{a+4z_lz_r}\right)}{4\left(a-\Delta_z^2\right)} \\
x^*&=\frac{a-\Delta_z\left(2 z_l+\sqrt{a+4 z_l z_r}\right)}{2\left(a-\Delta_z^2\right)}
\end{aligned}\label{xmax-zmax}
\end{eqnarray}
where $\Delta_z= z_l- z_r$.

\subsection{Global current rate function}
In addition to the ODP, we can calculate exactly the RF of the ZRP. This means solving Eq.~(\ref{macro-rf-simpl}) constrained by Eq.~(\ref{ellipse-equation-compact}) which, by the symmetry of our system reduces to calculating the integral
\begin{equation}
\hat{e}\left(\mb{J}\right)=\int\limits_{0}^{1}\left\{ \frac{\left(J_x+z'(\rho)p_x \frac{\partial\rho}{\partial x}\right)^2}{4p_xz(\rho)}+\frac{J_y^2}{4p_yz(\rho)}\right\}dx , \label{rf-macrozrp}
\end{equation}
with the minimizing constraint~(\ref{optprof}). Similarly to the procedure above, we use Eq.~(\ref{optprof1}) to work in terms of the fugacity (instead of the density), and substitute in Eq.~(\ref{rf-macrozrp}) to calculate the RF. Note that we still have to account for the change of regime for currents larger than the threshold given in~(\ref{transitioncurr-nonint}). This means that, in the monotonous scenario, the RF is obtained by integrating
\begin{equation}
\hat{e}(\mb{J})=-\int\limits_{ z_l}^{ z_r}\left\{\frac{p_x \left(a+\frac{2zC}{p_x}\right)}{2z\sqrt{a+\frac{4zC}{p_x}}}-\frac{J_x}{2z}\right\}d z, \label{rf-monot}
\end{equation}
while, for the non-monotonous regime, it becomes
\begin{equation}
\begin{aligned}
\hat{e}(\mb{J})&=\int\limits_{ z_l}^{ z^*} \left\{ \frac{	p_x a \left( 1 -\frac{z }{2z^*} \right) }{	2z \sqrt{ a\left( 1-\frac{z }{z^*} \right) }	}+\frac{J_x }{2z }\right\}d z \\
&- \int\limits_{ z^*}^{ z_r} \left\{ \frac{	p_x a \left( 1 -\frac{z }{2z^*} \right) }{	2z \sqrt{ a\left( 1-\frac{z }{z^*} \right) }	}-\frac{J_x}{2z }\right\} dz
\end{aligned},\label{rf-nonmonot}
\end{equation}
where the constants $C$ and $ z^*$ are determined by~(\ref{intconstant}) and~(\ref{xmax-zmax}). The exact solution of the integrals~(\ref{rf-monot}) and~(\ref{rf-nonmonot}) is given by
\begin{equation}
\begin{aligned}
\hat{e}(\mb{J}) &= \frac{J_x}{2}\ln\left(\frac{ z_r}{ z_l}\right)+\frac{p_x}{2}\left( \sqrt{a+b z_l} + \sqrt{a+b z_r} \right. \\
&- \left. 2\sqrt{a}\sinh^{-1}\left(\frac{a}{b z_l}\right) -2\sqrt{a}\sinh^{-1}\left(\frac{a}{b z_r}\right) \right) 
\end{aligned} \label{rf-nonint-monot}
\end{equation}
for the monotonous regime, and
\begin{equation}
\begin{aligned}
\hat{e}(\mb{J}) &= \frac{J_x}{2}\ln\left(\frac{ z_r}{ z_l}\right) + p_x\sqrt{a}\left[ \frac{\sqrt{1-\frac{ z_l}{ z^*}}+\sqrt{1-\frac{ z_r}{ z^*}} }{2} \right. \\
&- \left. \ln\left(\frac{\sqrt{ z_l z_r} \left( 1-\sqrt{1-\frac{ z_l}{ z^*}} \right) \left( 1-\sqrt{1-\frac{ z_r}{ z^*}} \right)}{ z^*}\right) \right]
\end{aligned}\label{rf-nonint-nonmonot}
\end{equation}
for the non-monotonous regime. For each specific ZRP interaction, the corresponding relation between density and fugacity can be used to obtain the RF in terms of the reservoir densities. In particular, for the case $w_n=n$ we can simply replace $z$ by $\rho$, whereas for $w_n=w$ we replace $z$ by $\rho/(\rho+1)$. Finally, to compare with the microscopic approach we have to relate the boundary fugacities to the boundary rates as mentioned in the main text.

\bibliography{masterbibmin}
\end{document}